\documentclass[a4paper,fleqn,usenatbib]{mnras}

\usepackage{newtxtext,newtxmath}

\usepackage[T1]{fontenc}
\usepackage{ae,aecompl}

\usepackage{graphicx}	
\usepackage{amsmath}	

\title[Small grains in the CGM]{Shattering as a source of small grains in the circum-galactic medium}

\author[H. Hirashita and T.-W. Lan]{
Hiroyuki Hirashita$^1$\thanks{E-mail: hirashita@asiaa.sinica.edu.tw}
and Ting-Wen Lan$^{2}$
\\
$^{1}$Institute of Astronomy and Astrophysics, Academia Sinica,
Astronomy-Mathematics Building,
No.\ 1, Sec.\ 4, Roosevelt Road, Taipei 10617, Taiwan\\
$^{2}$Department of Astronomy and Astrophysics, UCO/Lick Observatory, University of California,
1156 High Street, Santa Cruz, CA 95064, USA
}

\date{Accepted XXX. Received YYY; in original form ZZZ}

\pubyear{2020}

\begin{document}
\label{firstpage}
\pagerange{\pageref{firstpage}--\pageref{lastpage}}
\maketitle

\begin{abstract}
Observed reddening in the circum-galactic medium (CGM) indicates a significant abundance
of small grains, of which the origin is still to be clarified. We examine a possible path of
small-grain production through shattering of pre-existing large grains in the CGM.
Possible sites where shattering occurs on a reasonable time-scale are
cool clumps with hydrogen number
density $n_\mathrm{H}\sim 0.1$ cm$^{-3}$ and gas temperature
$T_\mathrm{gas}\sim 10^4$ K, which are shown to exist through observations of
Mg \textsc{ii} absorbers. We calculate the evolution of grain size distribution in physical
conditions appropriate for cool clumps in the CGM, starting from a large-grain-dominated
distribution suggested from theoretical studies.
With an appropriate gas turbulence model expected from the physical condition of cold clumps
(maximum eddy size and velocity of $\sim$100 pc and 10 km s$^{-1}$, respectively),
together with the above gas density and temperature
{and the dust-to-gas mass ratio inferred from
observations (0.006)},
we find that small-grain production occurs on a time-scale (a few $\times 10^8$ yr)
comparable to the lifetime of cool clumps derived in the literature.
Thus, the physical conditions of the cool clouds are favrourable
for small-grain production.
We also confirm that the reddening becomes significant on the above time-scale.
Therefore, we conclude that small-grain production by
shattering is a probable cause for the observed reddening in the CGM.
{We also mention the effect of grain materials (or their mixtures) on the
reddening at different redshifts (1 and 2).}
\end{abstract}


\section{Introduction}\label{sec:intro}

Dust grains form and evolve in galaxies through dust condensation in stellar
ejecta and various processes in the interstellar medium (ISM).
Dust has been not only observed in galaxies, but also found in the intergalactic
medium (IGM) or in the circum-galactic medium (CGM).
Background quasi-stellar objects (QSOs) are used
to trace the extinction properties of intervening absorption
systems distributed in a wide area of the Universe. This is made possible by
the availability of large statistical data such as those taken by
the Sloan Digital Sky Survey \citep[SDSS;][]{York:2000aa}.
\citet{York:2006aa} succeeded in deriving the dust extinction curves
of QSO absorption systems by comparing QSOs with and without intervening absorbers.

Using background QSOs, the dust properties in the CGM have been revealed observationally.
\citet[][hereafter M10]{Menard:2010aa} detected reddening (or colour excess, which is
caused by dust extinction) on a large scale (several Mpc) around galaxies
using the cross-correlation between the galaxy
position and the reddening of background QSOs for SDSS data, with median
$z\sim 0.3$ ($z$ denotes the redshift). Here the reddening is defined as the
difference in the extinction at two wavelengths.
\citet{Peek:2015aa} adopted basically the same method to
nearer galaxies ($z\sim 0.05$), and found a radial reddening profile similar to the
one found by M10.
\citet[][hereafter MF12]{Menard:2012aa} statistically detected reddening in Mg \textsc{ii} absorbers,
which trace the gas located in the CGM
\citep{Steidel:1994aa,Chen:2010aa,Nielsen:2013aa,Lan:2020aa},
confirming that the CGM contains dust.
\citet{Masaki:2012aa} used their analytic galaxy halo model
and reproduced the observationally suggested large extent of dust distribution in galaxy haloes.
Dust in the CGM is important in the total dust budget in the Universe,
since the dust mass in a galaxy halo is on average comparable to that in a galaxy disc
(M10; \citealt{Fukugita:2011aa}).

The dust in the CGM is in particular important since it is the interface between the galaxies
in which dust actually forms and the IGM which occupies a large volume of the Universe.
Thus, by investigating the dust in the CGM, we could clarify the mechanism with which
dust is dispersed in a wide region of the Universe.
There have been some efforts of clarifying the origin of the dust in the CGM.
Dust could be supplied into the CGM through galactic outflows driven by supernovae
(stellar feedback) and active galactic nuclei (AGNs) \citep[e.g.][]{Veilleux:2005aa}.
Some simulations indeed showed that galactic outflows transport the interstellar dust to the CGM
\citep{Zu:2011aa,McKinnon:2016aa,Hou:2016aa,Aoyama:2018aa}.
Radiation pressure from stars also push the interstellar dust outwards and
supply it to the CGM \citep{Ferrara:1991aa,Bianchi:2005aa,Bekki:2015aa,Hirashita:2019ab}.
These physical mechanisms could explain the existence of dust in the CGM.

In addition to reddening, dust in the CGM could also have a large influence
on the thermal evolution of the CGM. If the dust is contained in gas
with $T_\mathrm{gas}\gtrsim 10^6$ K ($T_\mathrm{gas}$ is the gas temperature), it radiates
away the thermal energy obtained through the collision with gas particles \citep[e.g.][]{Dwek:1987aa}.
In cooler gas, photoelectric heating by dust could play an important role in determining
the gas temperature
if the dust is irradiated by ultraviolet (UV) radiation \citep{Inoue:2003ab,Inoue:2004aa}.
Not only the dust abundance but also the grain size is important for the wavelength dependence of reddening
\citep[][hereafter HL20]{Hirashita:2020ab} and for the efficiency of photoelectric electron
emission \citep{Inoue:2003ab}. Therefore, to reveal the entire evolutionary history of the CGM dust,
it is crucial to clarify not only the total dust abundance but also the grain size
(or the grain size distribution) in the CGM.

The aforementioned mechanisms of grain injection from a galaxy to the CGM predict
typical sizes of transported grains. \citet{Hou:2017aa} showed, using a hydrodynamic simulation of
an isolated galaxy, that galactic winds driven by stellar feedback supply large
($a>0.03~\micron$, where $a$ is the grain radius) grains to the halo.
This is because dust grains formed in stellar ejecta are large
and are transported before they are significantly shattered in the ISM.
\citet{Aoyama:2018aa} confirmed the dominance of large grains in galaxy halos
using a cosmological simulation.
Another mechanism of grain transport, radiation pressure, also predicts a selective transport
of large ($a\sim 0.1~\micron$) grains.
\citet{Davies:1998aa} considered the motion of a grain in the gravitational potential
and the radiation field typical of a disc galaxy, and showed that dust grains with $a\sim 0.1~\micron$
can be ejected from the galactic disc.
According to their results, small grains with $a<0.01~\micron$ are not efficiently
transported because they are too small (compared with the wavelengths)
to receive radiation pressure efficiently.
\citet{Ferrara:1991aa} investigated the dust motion driven by radiation force towards
the halo in physical conditions typical of nearby spiral galaxies.
They showed that grain velocities could exceed 100 km s$^{-1}$
\citep[see also][]{Shustov:1995aa}, and that grains with $a\sim 0.1~\micron$ survive
against sputtering in the hot halo. \citet{Bianchi:2005aa},
by post-processing a cosmological simulation result at $z\sim 3$,
argued that large ($a\sim 0.1~\micron$) grains are preferentially injected into
the IGM, since smaller grains are decelerated by gas drag in denser regions near galaxies.
\citet{Hirashita:2019ab} also considered dust injection into galaxy halos focusing on
high-redshift galaxies, and showed that dust grains with $a\sim 0.1~\micron$ are the
most easily transported by radiation force. To summarize the above results,
both hydrodynamic galactic winds and radiation force selectively transport relatively large
($a\sim 0.1~\micron$) dust grains to galaxy halos.

HL20 showed, based on the reddening curves (reddening as a function of wavelength)
of objects tracing the CGM
(such as Mg\,\textsc{ii} absorbers; MF12), that galaxy haloes at $z\sim 1$--2 contain small dust grains
with $a\sim 0.01$--0.03 $\micron$. However, as mentioned above, such small grains
are not efficiently supplied to galaxy halos. This means that the origin of the small grains
indicated by the reddening observations is not clarified yet.

Shattering is a unique mechanism of producing small grains through the grain disruption
in grain--grain collisions. Although it is not likely that shattering is efficient
under the mean CGM density because of too low a number density of dust grains,
inhomogeneous structures in the CGM could enhance
the shattering efficiency locally. Indeed, it is suggested that the CGM contains
small \textit{cool} clumps as traced by Mg \textsc{ii} absorbers.
In this paper, the word `cool' is used to indicate that the temperature
is lower than the diffuse
surrounding medium. The typical temperature of cool gas is $T_\mathrm{gas}\sim 10^4$--$10^5$ K
(where $T_\mathrm{gas}$ is the gas temperature)
in this paper.
\citet{Lan:2017aa} analyzed relative strengths of various metal lines in
Mg \textsc{ii} absorbers and derived their typical gas density as $n_\mathrm{H}\sim 0.3$ cm$^{-3}$
(where $n_\mathrm{H}$ is the hydrogen number density).
Combining the density with the typical H \textsc{i} column density, they estimated
the typical dimension of Mg \textsc{ii} absorbers to be 30 pc.
Halo gas clumps of similar size have also been observed in our Milky Way halo
\citep[e.g.][]{BenBekhti:2009aa}
as well as in the CGM (more precisely the medium traced by QSO absorption lines)
at high redshift \citep[e.g.][]{Rauch:1999aa,Prochaska:2009aa}.

From the theoretical point of view,  such clumpy structures in the CGM could be produced
by shock compression and/or thermal instability, which is likely to induce
turbulent gas motion \citep{Buie:2018aa,Buie:2020aa,Liang:2020aa}.
The motions induced by galactic winds could also produce multiphase gas
\citep{Thompson:2016aa}. In this sense, turbulent motion in the CGM could be maintained
by the energy input from the central galaxy.
At the same time, if cool clumps coexist with hot diffuse gas, turbulent motion
is induced by creation of
rapidly cooled intermediate-temperature phase \citep{Fielding:2020aa}.
A velocity dispersion among dust grains also emerges in turbulence
\citep{Kusaka:1970aa,Volk:1980aa}, which could produce a
favourable condition for grain--grain collisions and shattering.
In general, an inhomogeneous density structure induced by turbulence
in the ISM has been shown to affect the dust evolution through enhanced
grain--gas or grain--grain collision rates
\citep{Mattsson:2020aa}. We expect that a similar argument
also holds for dust in the IGM.

The cool clumps may be transient structures, but their lifetimes could be
as long as $\sim\mbox{a few}\times 10^8$ yr.
\citet{Lan:2019aa} analytically estimated the evaporation time of clouds in pressure
equilibrium with the surrounding hot tenuous gas and found that the lifetime is
a few $\times 10^8$ yr for a cloud with a mass of $\sim 10^3$ M$_{\sun}$ appropriate for the
above clumps. This time-scale is also supported by hydrodynamic simulations
\citep{Armillotta:2017aa}, although the formation and destruction of cool clumps
can be affected
by the detailed treatment of thermal conduction and magnetic field
\citep{Sparre:2020aa,Li:2020aa}.

The turbulent motion on a spatial scale as small as 30 pc cannot be resolved in cosmological
or galaxy-scale simulations. Therefore, the above mentioned hydrodynamic simulations that
implemented the evolution of grain sizes were not able to treat
possible local enhancement of shattering associated with the CGM clumps. This means that
shattering in the CGM, even if it occurs, has been missed in previous works. Thus,
it is desirable to investigate if shattering in cool clumps really acts as a source of
small grains in the CGM.

The goal of this paper is to examine if shattering in the CGM is efficient enough to
produce small grains. We adopt some simple assumptions on the turbulent motion
associated with a cool clump in the CGM and calculate the evolution of
grain size distribution based on the grain--grain collision rate expected from the grain velocities
induced by the turbulence.
We also calculate reddening curves based on
the grain size distributions in order to test the consistency with
the observed reddening curves in the CGM (or in Mg\,\textsc{ii} absorbers).

This paper is organized as follows.
In Section~\ref{sec:model}, we describe the dust evolution model applied to
cool clumps in the CGM.
In Section~\ref{sec:result}, we show the results for the grain size distributions and
the reddening curves.
In Section \ref{sec:discussion}, we show some extended results and provide some
additional discussions including
prospects for future modelling.
In Section \ref{sec:conclusion}, we give the conclusion of this paper.

\section{Model}\label{sec:model}

For small-grain production by shattering in the CGM, grain velocities are important
in determining the impact and frequency of grain--grain collisions.
Grain velocities are sustained by turbulence.
Thus, in addition to the equation that describes shattering, we also explain how
to model the grain velocities driven by turbulence.
{The calculation of shattering only treats a single dust species because of
the difficulty in treating collisions among multiple species. We expect that as long
as the total grain abundance is correctly traced, the grain--grain collision rate is not
significantly under- or overestimated. In calculating the reddening later, we apply optical
properties of different dust species since the reddening is sensitive to the adopted
grain species (see Section \ref{subsec:param} for more descriptions for the treatment of
grain species).}

\subsection{Turbulent cool gas in the CGM}\label{subsec:turbulence}

As mentioned in the Introduction, we consider the multi-phase CGM,
in which cool clumps
($T_\mathrm{gas}\sim 10^4$--$10^5$~K; \citealt{Werk:2013aa})
coexist with the diffuse hot gas ($T_\mathrm{gas}\gtrsim 10^6$~K).
We also assume that, as expected from theoretical studies (see the Introduction),
turbulent motion associated with the cool clumps emerges. However,
the properties of the turbulence in the CGM are not fully understood.
Since the
energy input from galaxies is generally related to phenomena on a sub-galactic
spatial scale (starbursts, AGNs, etc.), we choose sub-kpc ($\sim 100$ pc)
for the injection scale of turbulence, $L_\mathrm{max}$.
This scale is also similar to the size of cool clumps in the CGM
\citep{Lan:2017aa}. We also assume that the turbulent velocity ($v_\mathrm{max}$)
at the maximum scale ($L_\mathrm{max}$)
in the cool gas
is comparable to the sound speed of the cool gas ($\sim 10$ km s$^{-1}$),
assuming that the gas temperature is broadly determined by Ly$\alpha$ cooling
(i.e.\ $T_\mathrm{gas}\sim 10^4$ K).

Under given $L_\mathrm{max}$ and $v_\mathrm{max}$,
we assume a Kolmogorov spectrum for the turbulence; that is, the typical velocity
$v$ on scale $\ell$ is written as
\begin{align}
v=v_\mathrm{max}\left(\frac{\ell}{L_\mathrm{max}}\right)^{1/3}.\label{eq:v}
\end{align}
We also define the turn-over time as
\begin{align}
\tau_\mathrm{turn}=\frac{\ell}{v}=\frac{L_\mathrm{max}v^2}{v_\mathrm{max}^3}.
\label{eq:tau_turn}
\end{align}
The turn-over time of the largest eddy is estimated as
$\tau_\mathrm{turn,max}=L_\mathrm{max}/v_\mathrm{max}$.

\subsection{Grain velocity}\label{subsec:vel}

In this paper, we assume grains to be spherical and compact,
so that $m=(4\upi /3)a^3s$, where $a$ is the grain radius and $s$
is the material density of dust.
The grain velocity is determined by the dynamical coupling with the turbulence
through gas drag. Since the mean free path of gas particles is much longer than the grain radius
in the situations treated in this paper, we apply the drag time-scale, $\tau_\mathrm{dr}$,
in the Epstein regime {\citep[e.g.][]{Weidenschilling:1977aa}}:
\begin{align}
\tau_\mathrm{dr} &= \frac{sa}{c_\mathrm{s}\rho_\mathrm{g}}\nonumber\\
&= 4.7\times 10^6\left(\frac{s}{3.5~\mathrm{g}~\mathrm{cm}^{-3}}\right)
\left(\frac{a}{0.1~\micron}\right)
\left(\frac{T_\mathrm{gas}}{10^4~\mathrm{K}}\right)^{-1/2}\nonumber\\
&\times \left(\frac{n_\mathrm{H}}{10^{-1}~\mathrm{cm}^{-3}}\right)^{-1}~\mathrm{yr},
\end{align}
where $c_\mathrm{s}$ is
the sound speed, and $\rho_\mathrm{g}$ is the gas density.
We applied $c_\mathrm{s}=10(T_\mathrm{gas}/10^4~\mathrm{K})^{1/2}$ km s$^{-1}$
(the precise value of this depends on the ionization state, but it only causes an uncertainty of
factor $\sim\sqrt{2}$)
and $\rho_\mathrm{g}=\mu n_\mathrm{H}m_\mathrm{H}$ ($\mu =1.4$ is the gas mass per hydrogen,
and $m_\mathrm{H}$ is the hydrogen atom mass).
This expression is applicable to subsonic grain motion, but it is approximately valid
even if the grain motion is slightly supersonic. Since we do not treat cases where grain motions are highly
supersonic, we use this expression for the drag time-scale.
For convenience, we also define the Stokes number, St, as
\begin{align}
\mathrm{St} &= \frac{\tau_\mathrm{dr}}{\tau_\mathrm{turn,max}}\nonumber\\
&= 0.49\left(\frac{s}{3.5~\mathrm{g}~\mathrm{cm}^{-3}}\right)
\left(\frac{a}{0.1~\micron}\right)
\left(\frac{T_\mathrm{gas}}{10^4~\mathrm{K}}\right)^{-1/2}\nonumber\\
&\times \left(\frac{n_\mathrm{H}}{10^{-1}~\mathrm{cm}^{-3}}\right)^{-1}
\left(\frac{L_\mathrm{max}}{100~\mathrm{pc}}\right)^{-1}
\left(\frac{v_\mathrm{max}}{10~\mathrm{km}~{s}^{-1}}\right) .
\label{eq:St}
\end{align}

{If the grain is small enough to satisfy $\mathrm{St}<1$
($\tau_\mathrm{dr}<\tau_\mathrm{turn,max}$),}
the grain velocity is determined by the maximum size of the eddies that the
grain can be coupled with: $\tau_\mathrm{dr}=\tau_\mathrm{turn}$.
This leads to the following estimate for the grain velocity (denoted as $v^\mathrm{(s)}$):
\begin{align}
v^\mathrm{(s)} &= v_\mathrm{max}^{3/2}\left(\frac{sa}{c_\mathrm{s}\rho_\mathrm{g}L_\mathrm{max}}\right)^{1/2}\nonumber\\
&= 7.0\left(\frac{s}{3.5~\mathrm{g}~\mathrm{cm}^{-3}}\right)^{1/2}\left(\frac{a}{0.1~\micron}\right)^{1/2}
\left(\frac{T_\mathrm{gas}}{10^4~\mathrm{K}}\right)^{-1/4}
\left(\frac{n_\mathrm{H}}{10^{-1}~\mathrm{cm}^{-3}}\right)^{-1/2}\nonumber\\
&\times\left(\frac{L_\mathrm{max}}{100~\mathrm{pc}}\right)^{-1/2}
\left(\frac{v_\mathrm{max}}{10~\mathrm{km}~{s}^{-1}}\right)^{3/2}~\mathrm{km~s}^{-1}.
\label{eq:vs}
\end{align}
Note that the above grain velocity is based on the same model as in \citet{Ormel:2009aa}
except that we treat $v_\mathrm{max}$ and $c_\mathrm{s}$ independently.

{In the large grain radius regime where $\mathrm{St}>1$
($\tau_\mathrm{dr}>\tau_\mathrm{turn,max}$), the grain is not coupled
even with the largest eddies.} The grain motion is
still fluctuated by the turbulence.
In this case, the grain velocity (denoted as $v^\mathrm{(l)}$) is estimated by
\citep{Ormel:2007ab}
\begin{align}
v^\mathrm{(l)}=\frac{v_\mathrm{max}}{\sqrt{1+\mathrm{St}}}.\label{eq:vl}
\end{align}
{Note that we use equation (\ref{eq:St}) for St and that $v^\mathrm{(l)}\propto a^{-1/2}$
at large grain radii.}

Since the boundary of the two regimes, $\tau_\mathrm{dr}\gtrless \tau_\mathrm{turn,max}$,
is equivalent to $\mathrm{St}=1$, the two cases can be roughly unified by setting the
grain velocity as
\begin{align}
v=\min (v^\mathrm{(l)},\, v^\mathrm{(s)}).
\end{align}
We adopt this expression for the grain velocity as a function of grain radius.

\subsection{Calculation of shattering}\label{subsec:shattering}

Based on the above grain velocities, we calculate the evolution of grain size distribution
by shattering. We basically follow \citet{Jones:1994aa,Jones:1996aa} and
\citet{Hirashita:2009ab}.
The grain size distribution at time $t$,
$n(a,\, t)$, is defined such that $n(a,\, t)\,\mathrm{d}a$ is the number density
of dust grains with grain radii between $a$ and $a+\mathrm{d}a$.
We also define the grain mass distribution,
$\varrho_\mathrm{d}(m,\, t)$, such that
$\varrho_\mathrm{d}(m,\, t)\,\mathrm{d}m$ is the mass density
of dust grains whose mass is between $m$ and $m+\mathrm{d}m$.
{Noting that the number density is converted to the mass density by multiplying
$m=(4/3)\upi a^3s$, we obtain}
\begin{align}
\varrho_\mathrm{d}(m,\, t)\,\mathrm{d}m=\frac{4}{3}\upi a^3sn(a)\,\mathrm{d}a,
\label{eq:rho_n}
\end{align}
with $\mathrm{d}m=4\upi a^2s\,\mathrm{d}a$.

The time evolution of grain mass distribution by shattering is expressed as
\citep{Hirashita:2019aa,Hirashita:2021aa}
\begin{align}
\lefteqn{\frac{\upartial\varrho_\mathrm{d} (m,\, t)}{\upartial t} = -m\varrho_\mathrm{d} (m,\, t)\int_0^\infty
\frac{K_{m,m_1}}{mm_1}\varrho_\mathrm{d} (m_1,\, t)\mathrm{d}m_1}\nonumber\\
& +\int_0^\infty\int_0^\infty\frac{K_{m_1,m_2}}{m_1m_2}\varrho_\mathrm{d} (m_1,\, t)
\varrho_\mathrm{d}(m_2,\, t)
m\bar{\theta} (m;\, m_1,\, m_2)\mathrm{d}m_1\mathrm{d}m_2,\label{eq:rho}
\end{align}
where $K$ is the collision kernel determined later (with the subscripts denoting the
masses of colliding grains),
$\tilde{\theta}(m;\, m_1,\, m_2)$,
{of which the functional form is explained below,}
describes the distribution function of
grains produced from $m_1$ in the collision between grains with masses $m_1$ and $m_2$.
The grain radii (or corresponding grain masses) are discretized into 128 grid points
in the range between 3 \AA\ and 10 $\micron$, and solve the discrete version of the
equation shown in \citet{Hirashita:2019aa}.
Note that we count the collision between
$m_1$ and $m_2$ twice and treat collisional products originating from
$m_1$ and $m_2$ separately. This is why we do not have a factor 1/2
(see equation A1 in \citealt{Jones:1994aa}) in front of the
second term on the
right-hand side in equation (\ref{eq:rho}).
When we consider collisions between grains with masses $m_1$ and
$m_2$ (radii $a_1$ and $a_2$, respectively), the kernel function is evaluated as
\begin{align}
K_{m_1,m_2}=\upi (a_1+a_2)^2\, v_{1,2}\, ,\label{eq:kernel}
\end{align}
where $v_{1,2}$ is the relative velocity between the
two grains.
In considering the collision between two grains
with $v=v_1$ and $v_2$, we estimate the relative velocity $v_{1,2}$ by
\begin{align}
v_{1,2}=\sqrt{v_1^2+v_2^2-2v_1v_2\mu_{1,2}}\,,\label{eq:rel_vel}
\end{align}
where $\mu_{1,2}=\cos\theta_{1,2}$ ($\theta_{1,2}$ is the angle between the two grain velocities)
is randomly chosen between $-1$ and 1 in every calculation of the kernel function
{(i.e.\ for each pair of grain bins at each time-step)}
\citep{Hirashita:2013aa}.

{The total shattered mass in the original grain $m_1$ is determined by the ratio
between the impact energy and the energy necessary for the catastrophic disruption
($Q_\mathrm{D}^\star$ per grain mass) following \citet{Kobayashi:2010aa}.
If the specific impact energy is much smaller than $Q_\mathrm{D}^\star$,
shattered mass ($m_\mathrm{ej}$) is negligible compared with the original grain mass ($m_1$);
if it is much larger than
$Q_\mathrm{D}^\star$, the whole grain is shattered ($m_\mathrm{ej}\simeq m_1$);
and if it is equal to $Q_\mathrm{D}^\star$,
half of the grain mass is shattered ($m_\mathrm{ej}=m_1/2$).
}

We basically take the mass distribution function of fragments,
{$\bar{\theta}(m;\, m_1,\, m_2)$,} from \citet{Hirashita:2019aa}
\citep[originally from][]{Kobayashi:2010aa}.
Briefly, $m_\mathrm{ej}$ is distributed into fragments, for which
we assume a power-law size distribution with an index of
$-3.3$ \citep{Jones:1996aa}, and the remnant $m_1-m_\mathrm{ej}$
is put in the appropriate grain radius bin.
The maximum and minimum masses of the fragments
are assumed to be
$m_\mathrm{f,max}=0.02m_\mathrm{ej}$ and
$m_\mathrm{f,min}=10^{-6}m_\mathrm{f,max}$, respectively \citep{Guillet:2011aa}
{that is, $\bar{\theta}=0$ if $m$ is not between $m_\mathrm{f,max}$
and $m_\mathrm{f,min}$.
We remove grains if the grain radius becomes smaller
than 3 \AA.}

{We neglect vaporization in this paper. As shown by \citet{Tielens:1994aa} and
\citet{Jones:1996aa}, shattering is dominant over vaporization in modifying the
abundance of both large and small grains even at collision velocities larger than 50 km s$^{-1}$.
In this paper, since we are interested in collision velocities $\lesssim 20$ km s$^{-1}$,
vaporization cannot be efficient than considered in the above papers. Thus,
neglecting vaporization does not affect the conclusion of this paper.}

\subsection{Calculation of reddening curves}\label{subsec:ext_method}

The wavelength dependence of dust extinction is used to test the small grain
production in the CGM. As done by HL20, we use
the colour excess or reddening, which is a relative flux change at two
wavelengths due to dust extinction.
The extinction at wavelength $\lambda$, $A_\lambda$, is estimated as
\begin{align}
A_\lambda =2.5(\log\mathrm{e})\,\kappa_\mathrm{ext}(\lambda )\mu m_\mathrm{H}
N_\mathrm{H}\mathcal{D} ,\label{eq:A_lambda}
\end{align}
where $\kappa_\mathrm{ext}(\lambda )$ is the mass extinction coefficient
(estimated below), $N_\mathrm{H}$ is the
column density of hydrogen nuclei, and $\mathcal{D}$ is the dust-to-gas mass ratio
{(hereafter, we simply refer $\mathcal{D}$ as the dust-to-gas ratio)}.
The dust-to-gas ratio is calculated using the following relation:
\begin{align}
\mu m_\mathrm{H}n_\mathrm{H}\mathcal{D}=\int_0^\infty\frac{4}{3}\upi a^3s\,
n(a,\, t)\,\mathrm{d}a.\label{eq:D}
\end{align}
Reddening, defined as the difference in the extinctions at two wavelengths
($\lambda$ and $\lambda_0$),
$A_\lambda -A_{\lambda_0}$,
is observationally derived from the background QSO colours
for Mg\,\textsc{ii} absorbers (MF12).
Thus, for the purpose of comparison, we calculate the reddening curve,
which is $A_\lambda -A_{\lambda_0}$ as a function
of $\lambda$.\footnote{Usually, we use extinction curve
$A_\lambda/A_{\lambda_0}$ \citep[e.g.][]{Pei:1992aa}, but the absolute value of
$A_{\lambda_0}$ is difficult to estimate for Mg \textsc{ii} absorbers.
Observationally, it is easier to
obtain the reddening from the method of MF12.}
The mass extinction coefficient
is estimated as
\begin{align}
\kappa_\mathrm{ext}(\lambda )=
\frac{\int_0^\infty\pi a^2Q_\mathrm{ext}(\lambda ,\, a)n(a)\,\mathrm{d}a}
{\int_0^\infty\frac{4}{3}\pi a^3sn(a)\,\mathrm{d}a}.\label{eq:kappa}
\end{align}
where $Q_\mathrm{ext}(\lambda,\, a)$ is the extinction cross-section
per geometric cross-section, and is calculated
using the Mie theory \citep{Bohren:1983aa} with
grain properties in the literature (Section \ref{subsec:param}).

The reddening is calculated by $A_\lambda -A_{i/(1+z)}$,
where $i$ is the $i$-band wavelength (0.76~$\micron$);
that is, we adopt $\lambda_0=i/(1+z)$.
Note that $\lambda$ is always used to indicate the rest-frame wavelength in this paper.
For observational data, we use the reddening curves for Mg \textsc{ii} absorbers at $z=1$
and 2 from MF12. We adopt the SDSS $u$, $g$, $r$, and $i$ band data, and do not use shorter
bands, where hydrogen, not dust, dominates the absorption (see fig.~4 in MF12).

Since the reddening is proportional to $N_\mathrm{H}$,
we need to specify the column density.
According to MF12, the typical column density of an Mg \textsc{ii} absorber
is $N_\mathrm{H}\sim 10^{19.5}$~cm$^{-2}$.
\citet{Lan:2017aa} also showed a range of $N_\mathrm{H}$ consistent with
the above, but with a larger scatter and redshift evolution.
\citet{Peek:2015aa} derived a reddening curve for galaxy halos at
$z\sim 0.05$. The SMC-like shapes of the reddening curves derived by these studies
are similar to that obtained from the comparison between
QSOs with and without absorption systems \citep{York:2006aa}.
Since the hydrogen column densities are better constrained for
Mg \textsc{ii} absorbers, we concentrate on the data in MF12
for comparison.

\subsection{Initial condition and parameters}\label{subsec:param}

Our model has the following unfixed parameters:
$n_\mathrm{H}$, $s$, $T_\mathrm{gas}$, $L_\mathrm{max}$, and $v_\mathrm{max}$.
The fiducial values and the ranges are summarized in Table~\ref{tab:param}.
\citet{Lan:2017aa} obtained $n_\mathrm{H}\sim 0.3~\mathrm{cm}^{-3}$
for Mg \textsc{ii} absorbers. We also expect that, if the typical density and temperature of
the CGM is $n_\mathrm{H}\sim 10^{-3}~\mathrm{cm}^{-3}$ and
$T_\mathrm{gas}\sim 10^6$~K, the pressure equilibrium
predicts that $n_\mathrm{H}\sim 10^{-1}$~cm$^{-3}$ for a cool
($T_\mathrm{gas}\sim 10^4$~K) gas \citep{Werk:2014aa}.
Thus, we adopt $n_\mathrm{H}=0.1$~cm$^{-3}$ for the fiducial value.
We also examine a less enhanced gas density, $n_\mathrm{H}=0.01$ cm$^{-3}$,
to demonstrate the importance of density enhancement for efficient shattering.
For the gas temperature, we adopt $T_\mathrm{gas}=10^4$ K, which is a typical temperature
achieved by atomic hydrogen cooling. We also examine $T_\mathrm{gas}=10^5$ K
(Section \ref{subsec:turbulence}). The maximum eddy size of turbulence, $L_\mathrm{max}$, is
assumed to be a sub-galactic scale ($\sim 100$ pc; Section \ref{subsec:turbulence}),
which is comparable to the clump size derived by \citet{Lan:2017aa}.
We also examine an order-of-magnitude range for $L_\mathrm{max}=30$--300 pc.
For $v_\mathrm{max}$, we adopt 10 km s$^{-1}$ for the fiducial value, assuming that the
turbulence velocity is of the same order of magnitude as the sound velocity. We also
examine $v_\mathrm{max}=5$--20 km s$^{-1}$ (based on the assumption that the
turbulent velocity is not very different from the characteristic sound speed).

\begin{table}
\caption{Parameters.}
\begin{center}
\begin{tabular}{lcccc}
\hline
Parameter & units & fiducial value & minimum & maximum \\
\hline
$n_\mathrm{H}$ & cm$^{-3}$ & 0.1 & 0.01 & 1 \\
$T_\mathrm{gas}$ & K & $10^4$ & $10^4$ & $10^5$\\
$L_\mathrm{max}$ & pc & 100 & 30 & 300\\
$v_\mathrm{max}$ & km s$^{-1}$ & 10 & 5 & 20 \\
$\mathcal{D}$ & & $0.006$ & \multicolumn{2}{c}{fixed}\\
\hline
\end{tabular}
\label{tab:param}
\end{center}
\end{table}

The dust supplied from the central galaxy to the IGM is dominated by large
grains based on the results of simulations and calculations mentioned in the
Introduction. Also, it is our central goal to investigate the possibility of producing
small grains from large grains supplied from the central galaxy. Thus,
we assume that the dust abundance is dominated by large grains
($a\sim 0.1~\micron$). For the initial grain size distribution, we adopt a simple
functional form described by the following lognormal function
with the {characteristic} grain size $a_0$ and the standard deviation $\sigma$:
\begin{align}
n(a,\, t=0)=\frac{C_0}{a}\exp\left\{ -
\frac{[\ln (a/a_0)]^2}{2\sigma^2}\right\} ,
\end{align}
where $C_0$ is the normalization constant determined below.
We adopt $a_0=0.1~\micron$ based on the above typical grain radius,
and $\sigma =0.47$ based on the grain size distribution
for the dust produced by stars in \citet{Asano:2013aa} and \citet{Hirashita:2019aa}.
The specific choice of $\sigma$ is not essential in this paper.
The normalization constant is determined by the initial dust-to-gas ratio,
$\mathcal{D}_0$, using equation~(\ref{eq:D}) with $\mathcal{D}=\mathcal{D}_0$
and $t=0$.
The typical dust-to-gas ratio of Mg\,\textsc{ii} absorbers is
60--80 per cent of the Milky Way value if we use
$A_V/N_\mathrm{H}$ for the indicator of dust-to-gas ratio (M12).
The initial dust-to-gas ratio $\mathcal{D}_0$ scales the
frequency of grain--grain collisions. Thus, we fix $\mathcal{D}_0=0.006$
(see also \citealt{Menard:2009aa}; HL20), keeping in mind
that the time-scale simply scales with $\mathcal{D}_0^{-1}$ (Section \ref{subsec:ext_method}).
Note that $\mathcal{D}$ decreases at later epochs because of the loss of dust through the
lower boundary of grain radius.

For the calculations of shattering, we adopt the silicate properties,
$s=3.5$ g cm$^{-3}$ and $Q^\star_\mathrm{D}=3\times 10^{11}$ dyn cm$^{-2}$
\citep{Jones:1996aa,Weingartner:2001aa,Hirashita:2013ab}
unless otherwise stated.
If we adopt the graphite properties instead, shattering proceeds faster
because of lower tensile strength (i.e.\ easier disruption) and
lower material density (i.e.\ more dust grains
under a fixed dust mass abundance). We confirmed that the time-scale of small-grain production
for graphite is
1/3 of that for silicate. Thus, adopting the silicate properties gives a conservative estimate for
the time required for small-grain production in the sense that the evolution could be three times faster
with graphite properties.

The grain species is rather essential to the reddening curve.
Thus, in calculating the reddening curve, we examine the diversity of grain species.
We adopt silicate and carbonaceous dust
for representative grain species.
Following HL20, we consider graphite and amorphous carbon (amC) for carbonaceous species.
We evaluate $\kappa_\mathrm{ext}$ (equation~\ref{eq:kappa})
based on the calculated grain size distribution $n(a)$
but using the grain material
densities for each grain material density $s$ (3.5, 2.24, and 1.81 g cm$^{-3}$ for
silicate, graphite, and amC, respectively; \citealt{Weingartner:2001aa,Zubko:2004aa}).
The extinction efficiency, $Q_\mathrm{ext}(\lambda ,\, a)$, is calculated using the
grain optical properties taken from \citet[][and references therein]{Weingartner:2001aa}
for silicate and graphite,
and from \citet{Zubko:1996aa} (`ACAR') for amC.
When we evaluate the dust-to-gas ratio $\mathcal{D}$ (equation \ref{eq:D}),
which is used in equation (\ref{eq:A_lambda}),
we always adopt the silicate
value ($s=3.5$ g cm$^{-3}$)
to achieve the same total dust mass abundance. In other words, our comparison among
grain species is fair in the sense that the underlying grain size distributions and
the total dust mass are the same.

\section{Results}\label{sec:result}

\subsection{Grain velocities}

\begin{figure}
\includegraphics[width=0.45\textwidth]{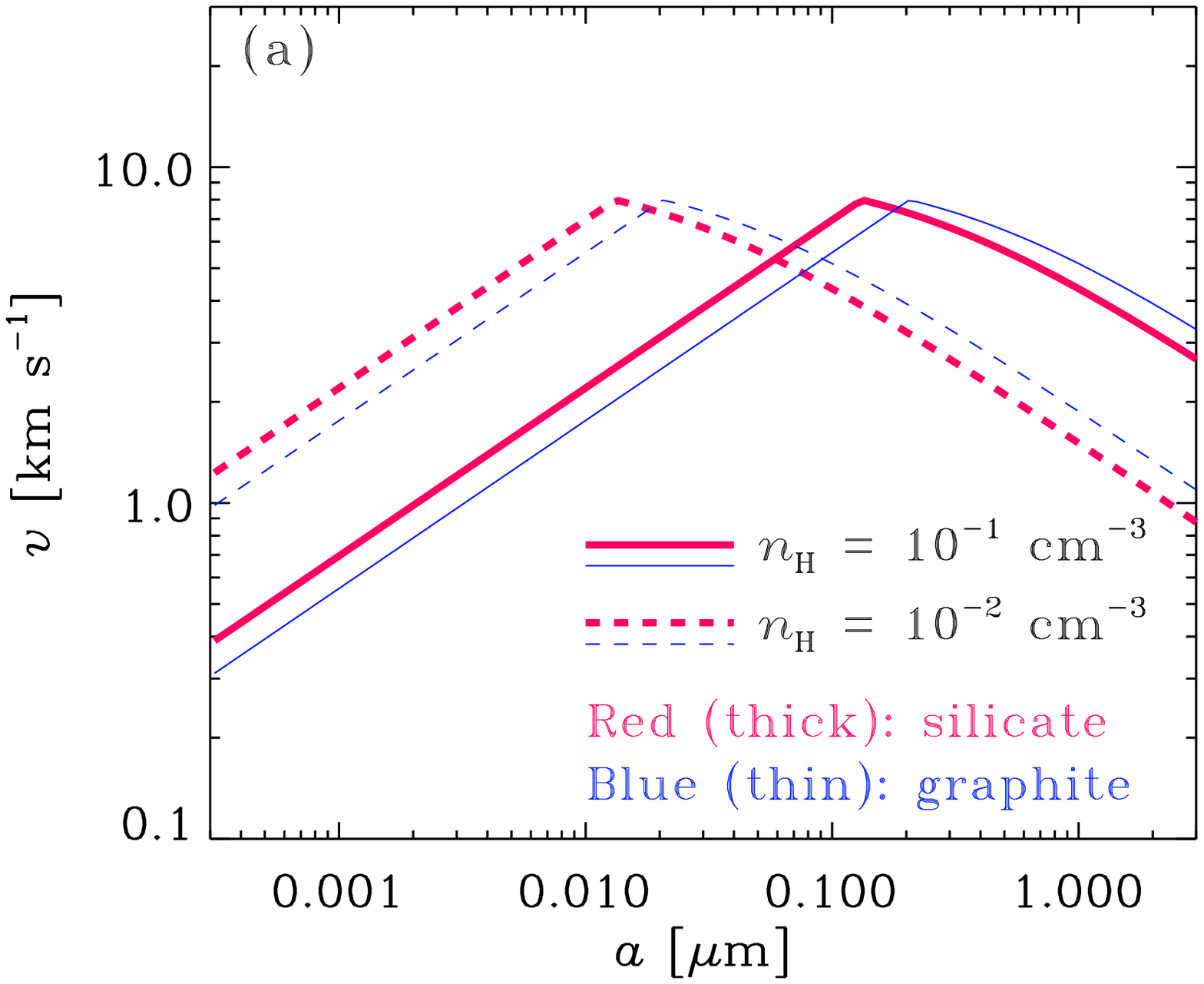}
\includegraphics[width=0.45\textwidth]{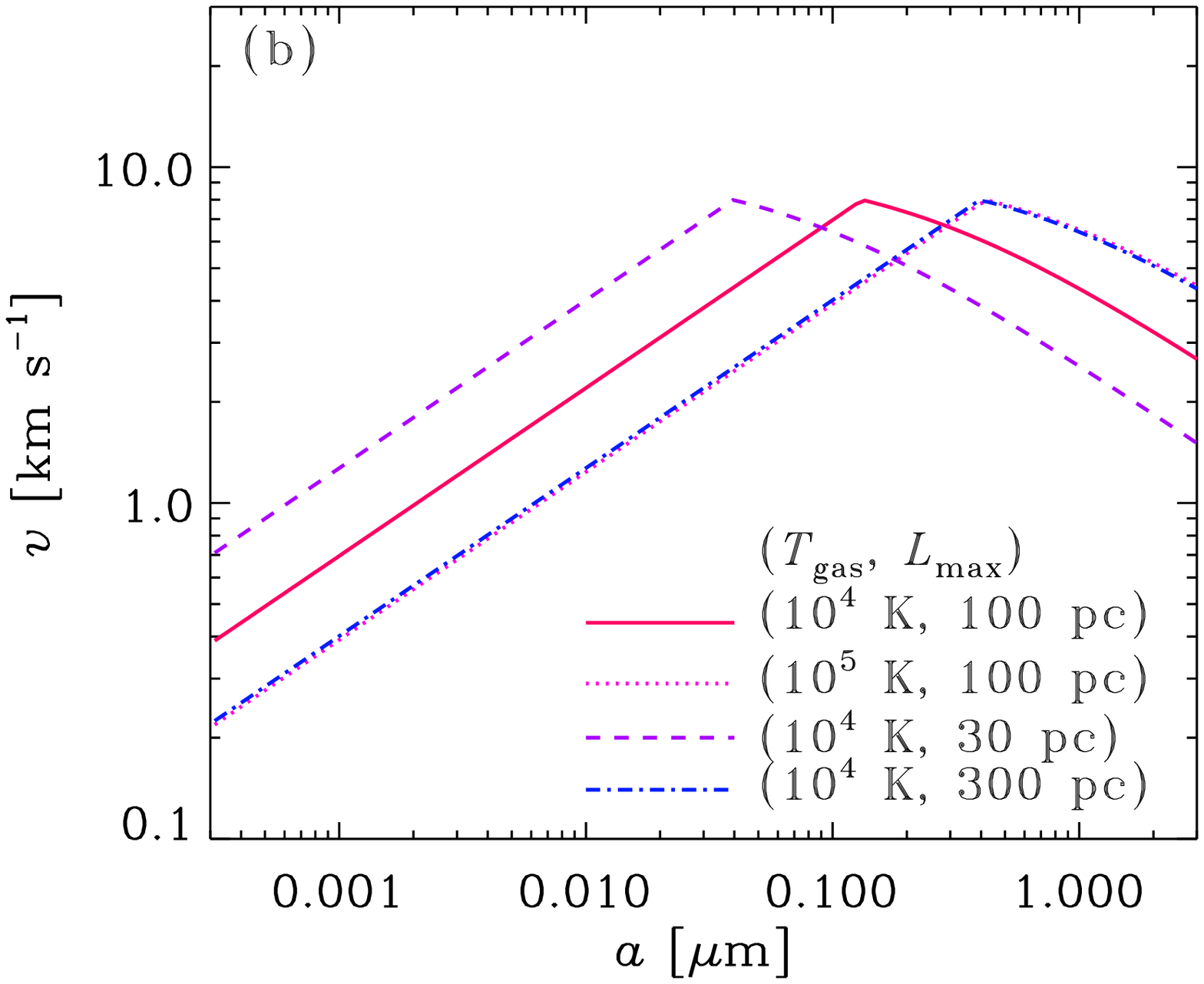}
\includegraphics[width=0.45\textwidth]{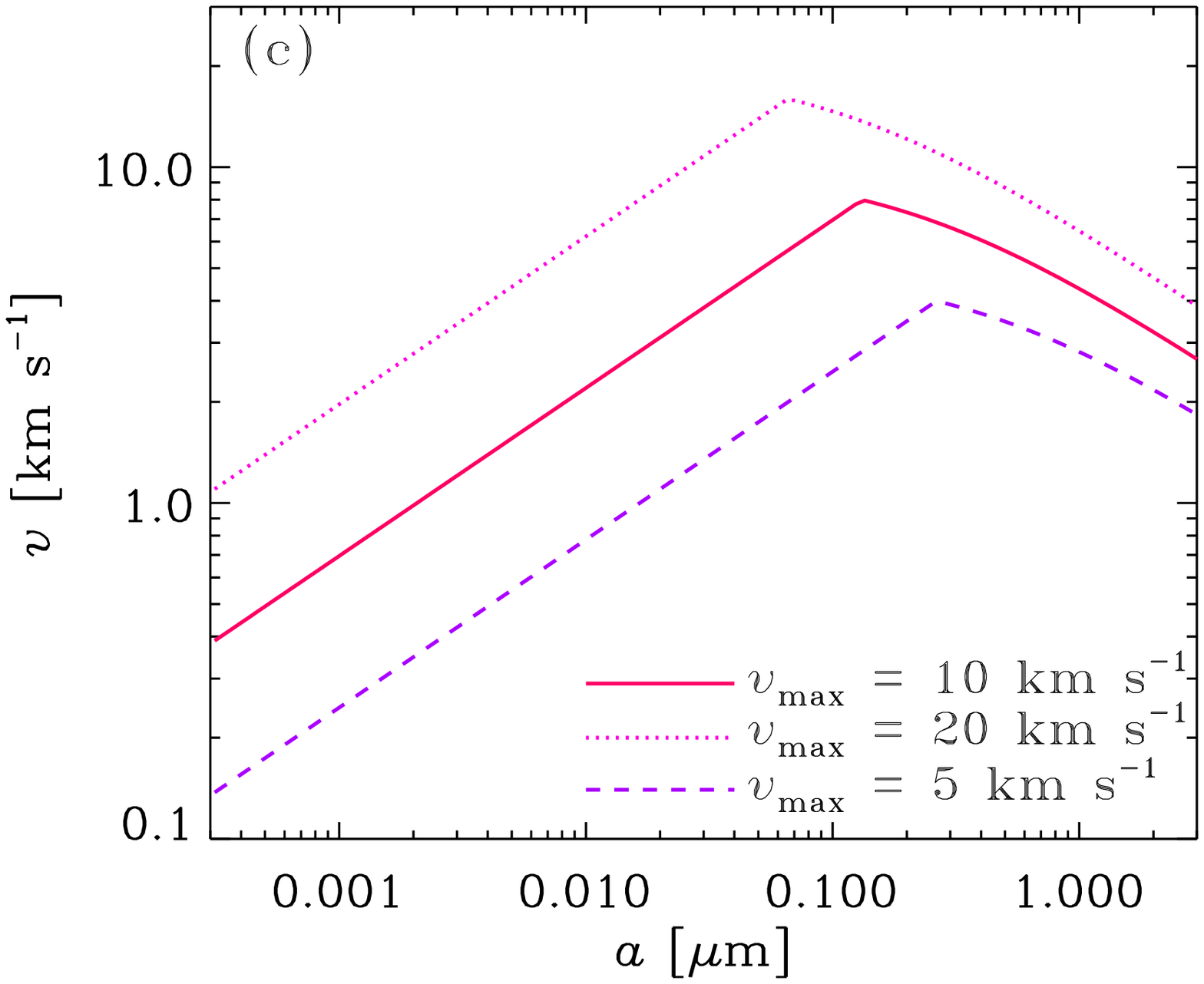}
\caption{Grain velocity as a function of grain radius $a$.
We adopt the fiducial values for the parameters but vary one of them.
(a) Dependence on hydrogen number density
$n_\mathrm{H}$. The solid and dotted lines present the cases with
$n_\mathrm{H}=0.1$ and 0.01 cm$^{-3}$, respectively. The thick (red) and thin (blue) lines
show the results for silicate and graphite, respectively (i.e.\ different grain material
densities). (b) Dependence on gas temperature $T_\mathrm{gas}$ and
maximum turbulence eddy size $L_\mathrm{max}$. The solid line shows the fiducial
case. The dotted line presents the case where the temperature is varied to
$10^5$ K (with $L_\mathrm{max}=100$ pc). The dashed and dot--dashed lines indicate
$L_\mathrm{max}=30$ and 300 pc, respectively (with $T_\mathrm{gas}=10^4$ K).
Note that the dotted and dot--dashed lines
almost overlap. (c) Dependence on the velocity of the largest eddy, $v_\mathrm{max}$.
The solid, dotted, and dashed lines present $v_\mathrm{max}=10$ (fiducial), 20, and 5 km s$^{-1}$,
respectively.
\label{fig:vel}}
\end{figure}

The impact and frequency of grain--grain collisions are affected by the grain velocity
calculated by the method in Section \ref{subsec:vel}.
Because the grain velocity is important in
interpreting the evolution of grain size distribution, it is useful to present it before we
show the main results.

We show the dependence of grain velocity $v$ on the parameters in Fig.\ \ref{fig:vel}.
Recall that $v$ is a function of grain radius $a$.
Overall, the grain velocity increases up to a certain grain radius and decreases at large $a$.
At small $a$, grains are coupled with the turbulence, so that the grain velocity follows
$v^\mathrm{(s)}$ (equation \ref{eq:vs}). At large $a$, grains are not fully coupled even with the
largest turbulence eddies, so that the grain velocity decreases with $a$
following $v^\mathrm{(l)}$
in equation (\ref{eq:vl}). The slopes at large and small $a$ are common for all cases.

In Fig.\ \ref{fig:vel}(a), we show the effect of gas density on the grain velocity.
If the gas is denser, the grain velocity is reduced at small $a$ because of stronger gas drag
(i.e.\ grains are coupled with smaller-scale motions, which have lower velocities).
On the other hand, the grain velocity is raised at large $a$ in denser environments because
the decoupling effect becomes weaker (i.e.\ the grain motion is more efficiently fluctuated by
the largest-scale turbulent motions). Thus, the gas density has an opposite
effect between small and large grains. In general, if the grains are more easily coupled dynamically
with the gas, the velocity at small $a$ decreases and that at large $a$ increases.
The peak velocity does not change as long as
$v_\mathrm{max}$ is fixed. In the same figure, we also examine the grain material dependence.
Graphite is more efficiently coupled with the turbulence than silicate because of its smaller inertia.
Therefore, graphite grains have lower velocities at small $a$ and higher velocities at large $a$
than silicate. However, the difference in the grain velocities between silicate and graphite is
small compared with that caused by other parameters. Thus, we use the grain velocities calculated
for silicate in this paper.

In Fig.\ \ref{fig:vel}(b), we present the dependence on $T_\mathrm{gas}$ and
$L_\mathrm{max}$. If $T_\mathrm{gas}$ is higher, the grain--gas coupling becomes
more efficient because of a higher collision rate between gas and dust; thus, the
grain velocities are lower at small $a$ and higher at large $a$.
On the other hand, if $L_\mathrm{max}$ is larger, the turbulent velocity is reduced
(equation \ref{eq:v}). Accordingly, the grain motion responds to the turbulent motion
more easily, which means that the grains are more efficiently coupled with the turbulence.
Thus, for larger $L_\mathrm{max}$, the grain velocities become lower at small $a$ and
higher at large $a$. Note that the same value of
$T_\mathrm{gas}L_\mathrm{max}^2$ produces the same result because of
the scaling in equation (\ref{eq:vs}).

In Fig.~\ref{fig:vel}(c), we examine the dependence on $v_\mathrm{max}$.
We observe that $v_\mathrm{max}$
affects the grain velocities at all grain radii in the same direction.
In addition, the grain radius where the grain velocity peaks shifts towards smaller $a$
for larger $v_\mathrm{max}$
since grains are less coupled with higher-velocity turbulent motions (owing to shorter
turn-over time-scales).
It is important to note that only $v_\mathrm{max}$ changes the overall grain velocity
level (or the peak grain velocity), while the other parameters investigated above only vary
the grain radius at which the grain velocity peaks.

\subsection{Evolution of grain size distribution}\label{subsec:size_result}

\begin{figure}
\includegraphics[width=0.45\textwidth]{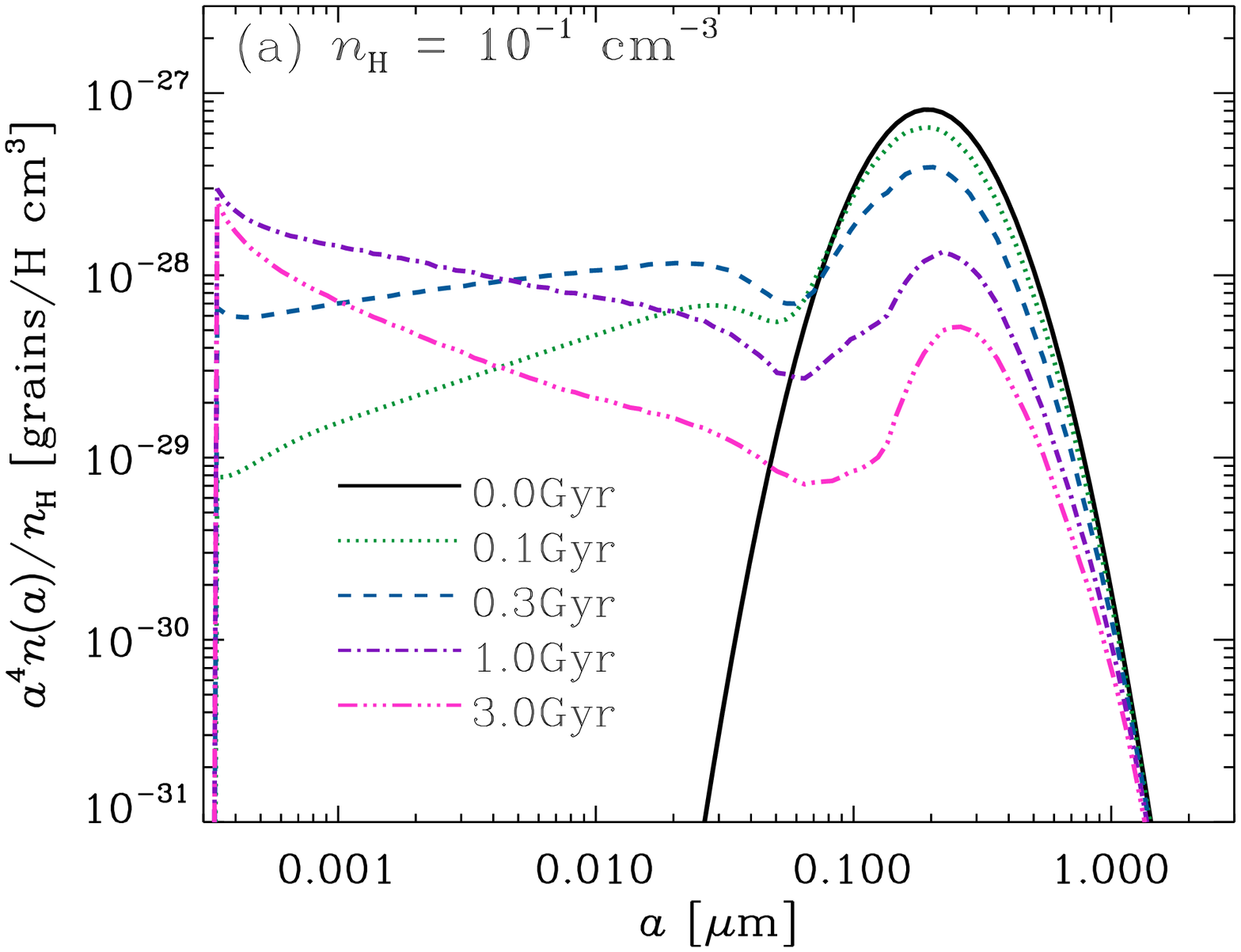}
\includegraphics[width=0.45\textwidth]{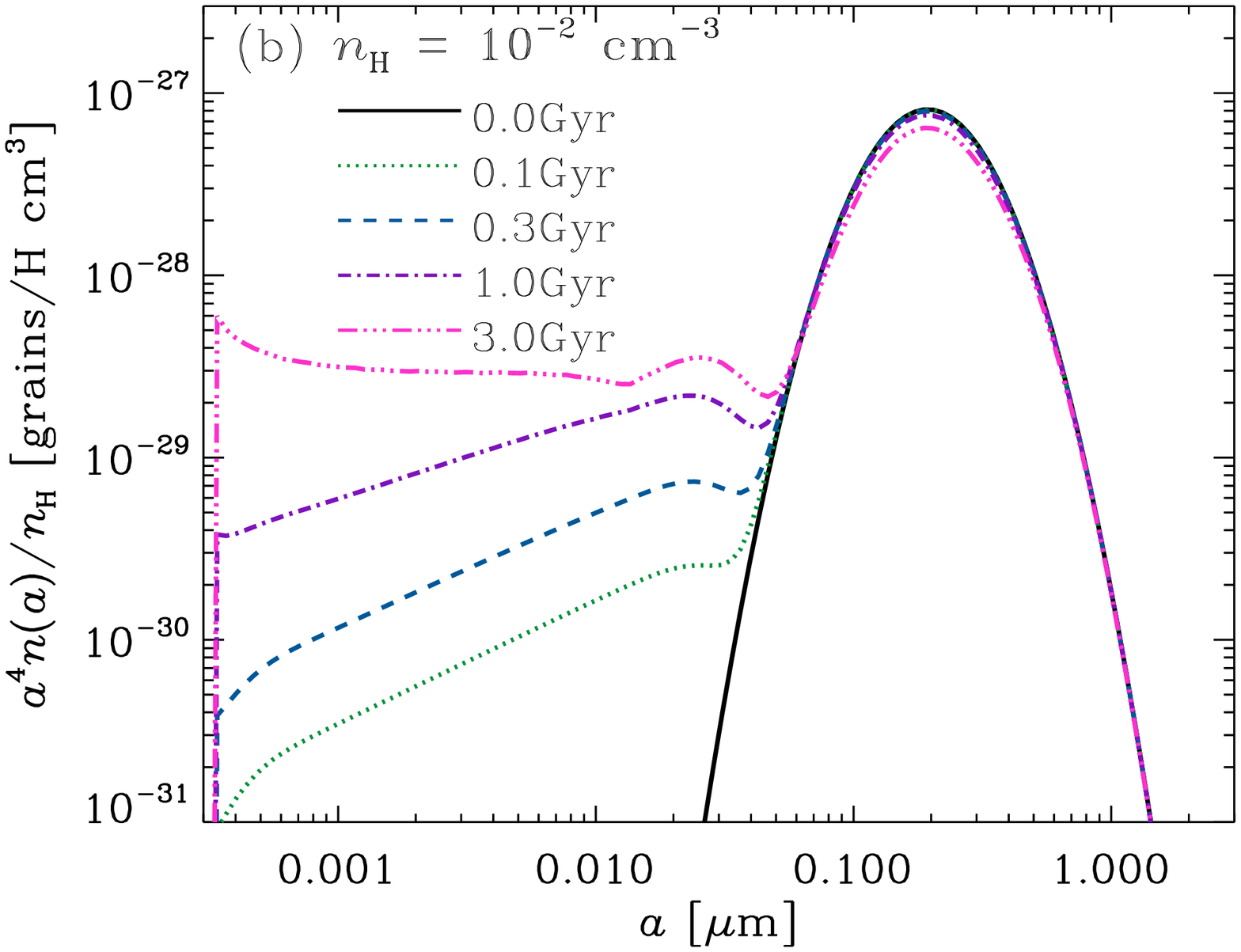}
\caption{Evolution of grain size distribution by shattering.
The solid, dotted, dashed, dot--dashed, and triple-dot--dashed lines show the results
at $t=0$ (initial condition), 0.1, 0.3, 1, and 3~Gyr, respectively.
We show the results for (a) $n_\mathrm{H}=10^{-1}$~cm$^{-3}$
and (b) $n_\mathrm{H}=10^{-2}$ cm$^{-3}$ with the other parameters fixed to the fiducial values
(Table \ref{tab:param}).
{Note that the vertical axis $a^4n(a)/n_\mathrm{H}$ is proportional
to the mass-weighted grain size distribution per $\log a$ relative to
the gas mass.
In this expression, the peak of the initial condition is
located at $a=a_0\exp (3\sigma^2)\simeq 0.19~\micron$.}
\label{fig:size}}
\end{figure}

Based on the above results for the grain velocities, we examine the evolution
of grain size distribution focusing on the dependences
on the following parameters: $n_\mathrm{H}$, $L_\mathrm{max}$, and
$v_\mathrm{max}$.
The gas temperature and the maximum turbulence eddy
size are degenerate in such a way that the same value of
$T_\mathrm{gas}L_\mathrm{max}^2$ predicts the same evolution of grain size distribution
as mentioned above. In other words, the case of $T_\mathrm{gas}=10^5$~K (with
$L_\mathrm{max}=100$~pc) can be
effectively examined by $L_\mathrm{max}\sim 30$ pc (with $T_\mathrm{gas}=10^4$~K).
Since $v_\mathrm{max}$ and $n_\mathrm{H}$ affect the grain--grain collision rate directly,
we examine the dependence on these quantities separately.
We move one parameter and fix the others to the fiducial values (Table \ref{tab:param}).

We show the evolution of grain size distribution in Fig.\ \ref{fig:size} for $n_\mathrm{H}=10^{-1}$
and $10^{-2}$~cm$^{-3}$. For $n_\mathrm{H}=10^{-1}$~cm$^{-3}$, we observe that an
appreciable fraction of large grains are shattered on a time-scale of a few $\times 10^8$ yr.
In contrast, in the case of $n_\mathrm{H}=10^{-2}$ cm$^{-3}$, the abundance of small grains is
much lower than that of large grains even at $t=3$ Gyr.
The gas density affects the grain size distribution in
the following two ways: one is that the grain--grain collision rate is proportional to
$n_\mathrm{H}$ (under a fixed dust-to-gas ratio),
and the other is that the grain velocity decreases at $a\sim 0.1~\micron$ for lower $n_\mathrm{H}$
because the grains are less coupled with the largest eddies (Fig.\ \ref{fig:vel}a).
Because of the second effect,
the shattering time-scale becomes further longer than expected from a simple scaling
$\propto n_\mathrm{H}^{-1}$ for lower densities.
Thus, the small-grain production rate by shattering is
very sensitive to the gas density.

\begin{figure}
\includegraphics[width=0.45\textwidth]{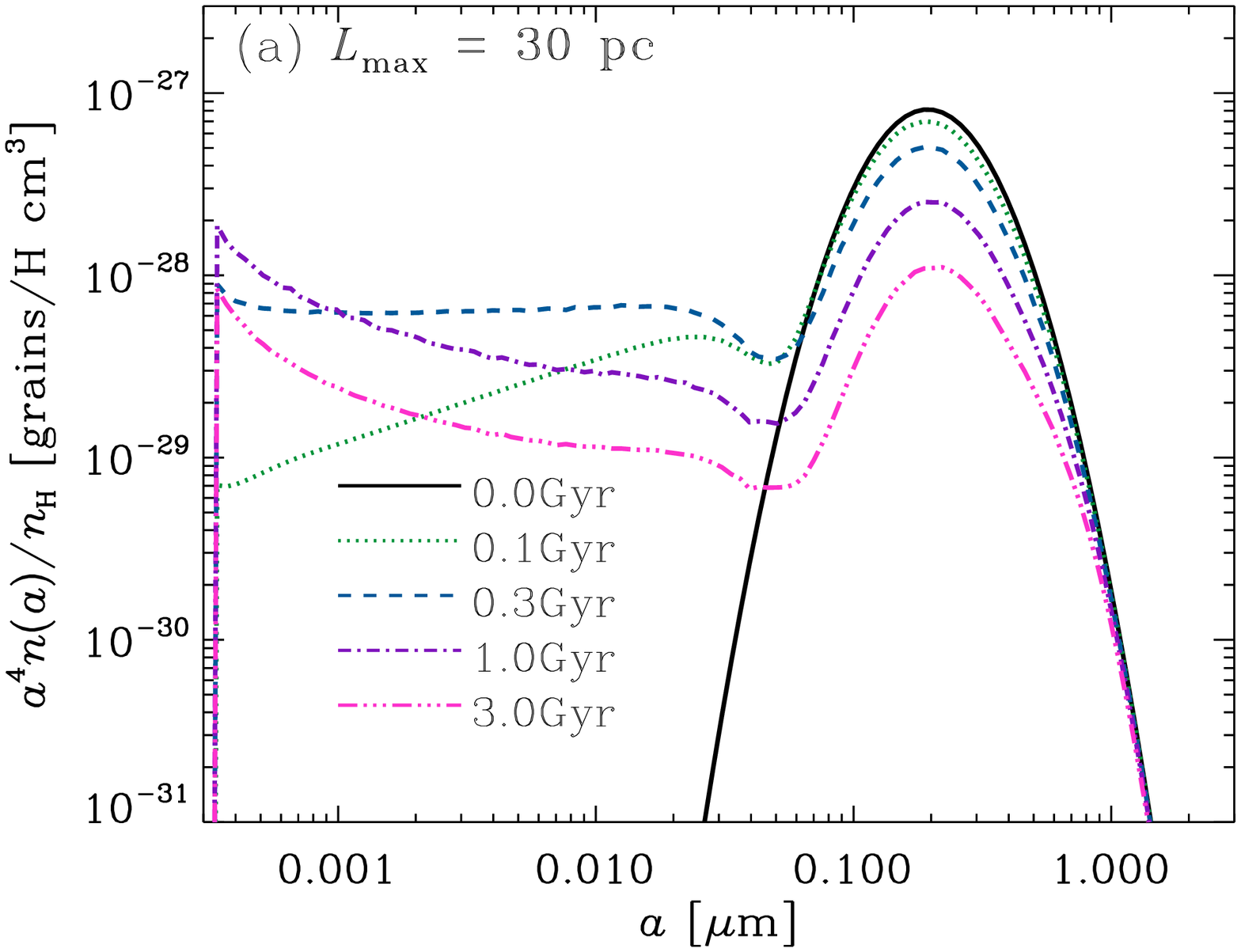}
\includegraphics[width=0.45\textwidth]{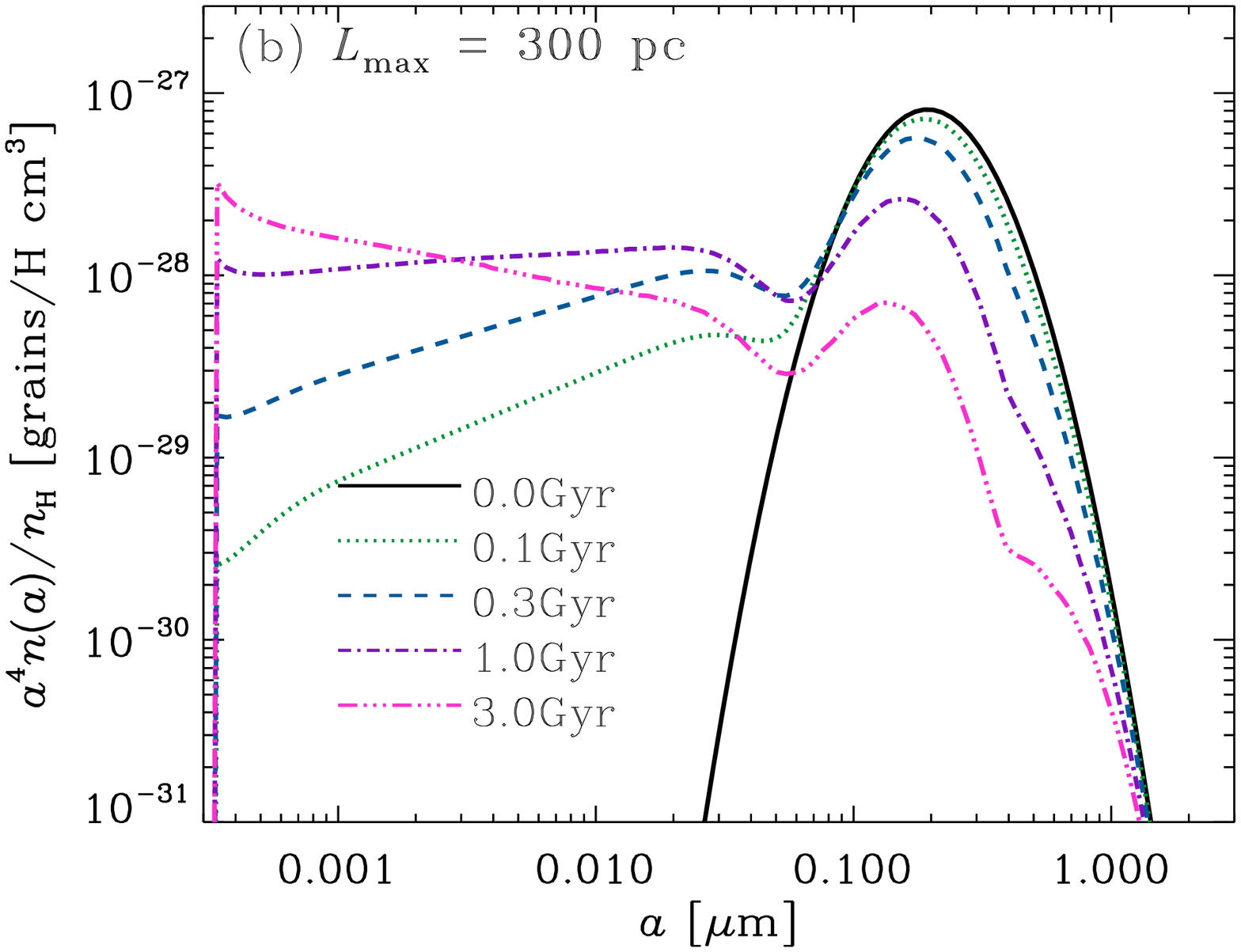}
\caption{Same as Fig.~\ref{fig:size}(a) but for $L_\mathrm{max}=30$ and 300~pc
in Panels (a) and (b), respectively.
For the other quantities, we adopt the fiducial values (Table~\ref{tab:param}).
Note that the case with $L_\mathrm{max}=100$~pc is shown in Fig.~\ref{fig:size}(a).
\label{fig:size_L}}
\end{figure}

Next, we examine the effect of $L_\mathrm{max}$ in Fig.~\ref{fig:size_L}.
At $t=0.3$ and 1 Gyr,
small-grain production for $L_\mathrm{max}=30$ and 300 pc is slightly
less efficient
compared with the results for $L_\mathrm{max}=100$ pc
(note that the case with $L_\mathrm{max}=100$~pc is shown in Fig.\ \ref{fig:size}a).
This is because the grain radius at which the grain velocity peaks is a little off
the peak of grain size distribution (Fig.\ \ref{fig:vel}b).
In other words,
$L_\mathrm{max}=100$ pc is optimum for shattering {if the characteristic grain
radius is $a_0\simeq 0.1~\micron$}.
This means that the energy input to turbulence on a spatial scale of
$\sim 100$ pc is favoured for small-grain production.
Note again that the same result is obtained
for the same value of $T_\mathrm{gas}L_\mathrm{max}^2$.
Thus, we can also conclude that $T_\mathrm{gas}=10^4$ K is
favoured for small-grain production {as long as $a_0\simeq 0.1~\micron$}.

\begin{figure}
\includegraphics[width=0.48\textwidth]{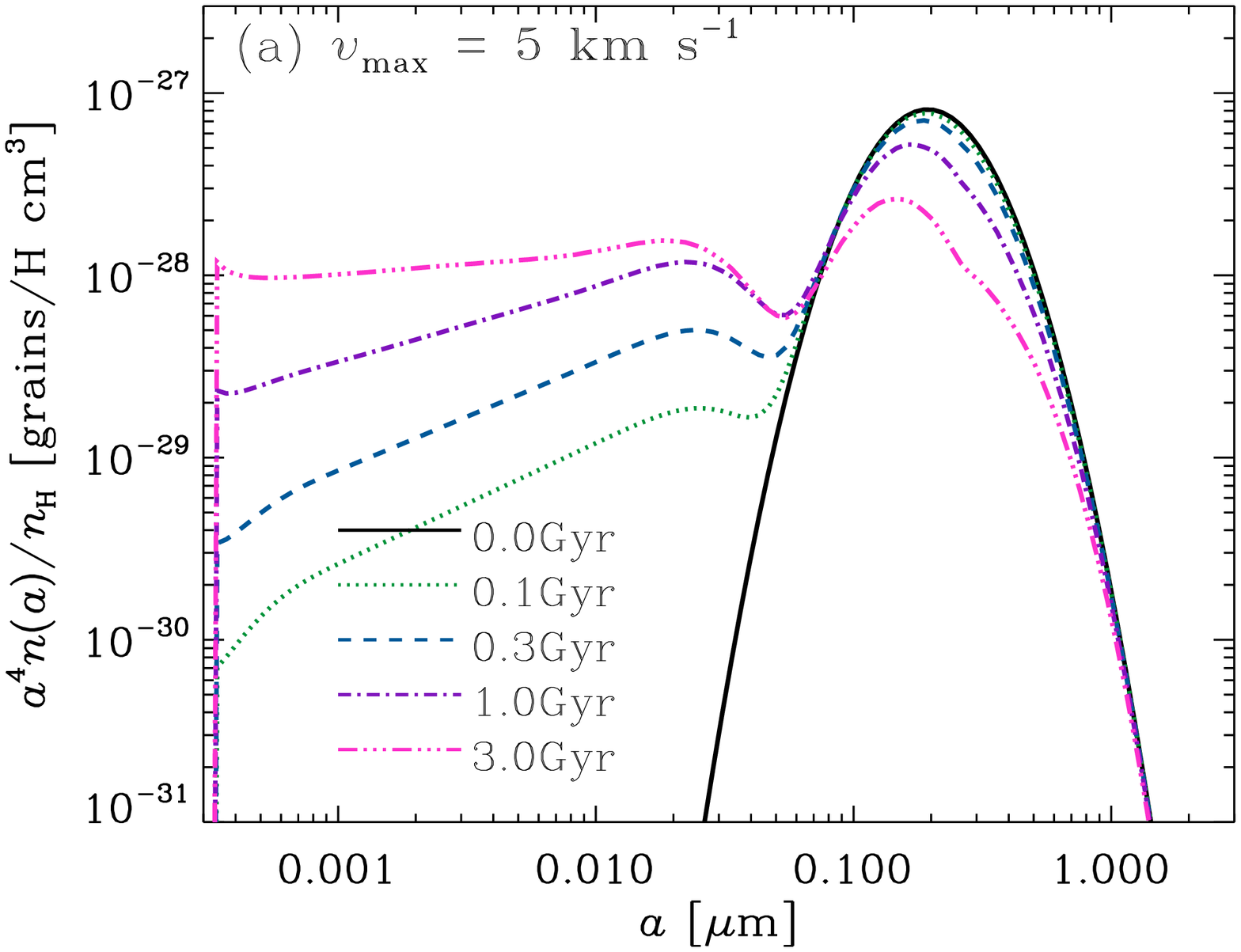}
\includegraphics[width=0.48\textwidth]{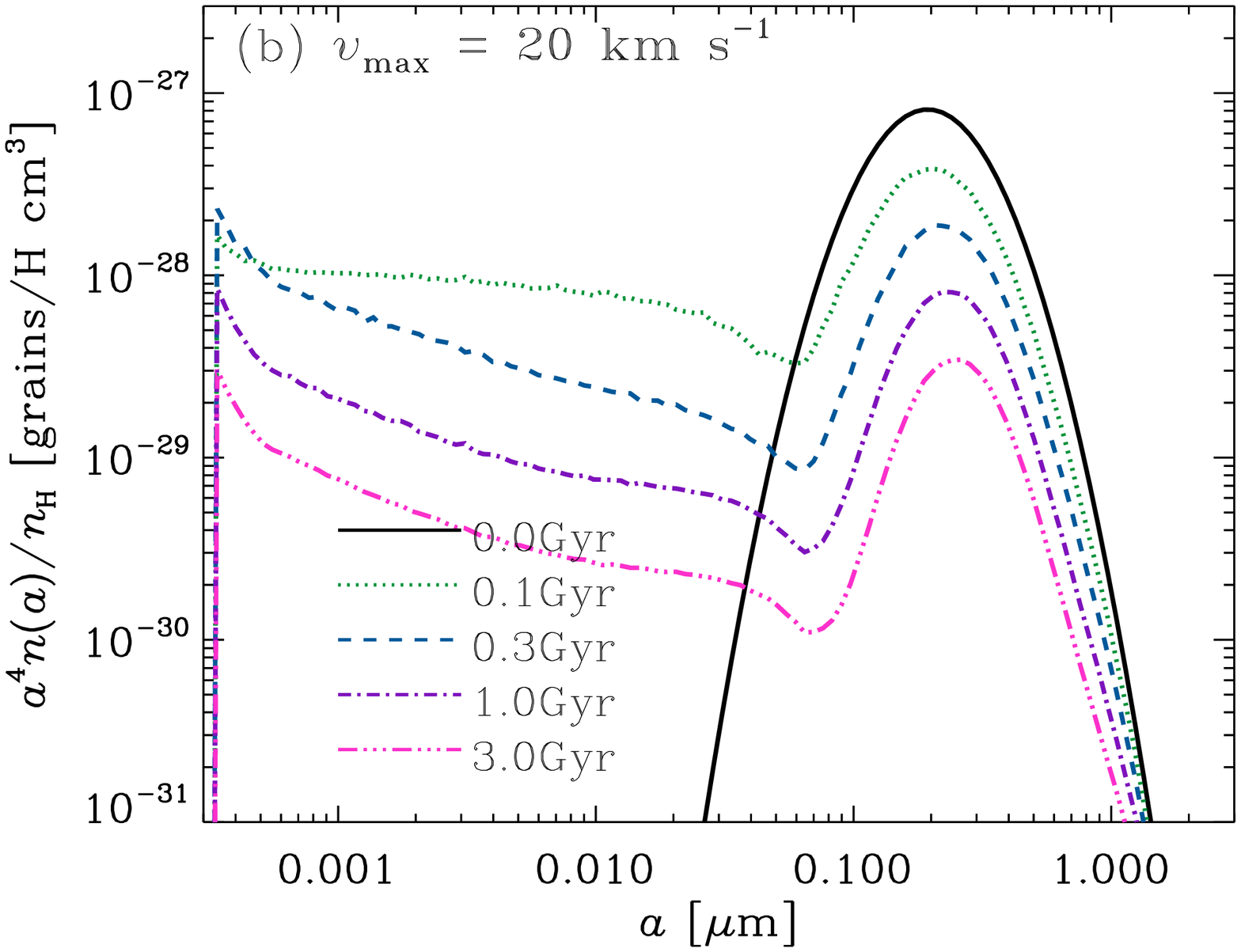}
\caption{Same as Fig.\ \ref{fig:size}(a) but for $v_\mathrm{max}=5$ and 20 km s$^{-1}$
in Panels (a) and (b), respectively.
For the other quantities, we adopt the fiducial values.
Note that the case with $v_\mathrm{max}=10$ km s$^{-1}$ is shown in Fig.\ \ref{fig:size}(a).
\label{fig:size_v}}
\end{figure}

Finally, we examine the effect of $v_\mathrm{max}$.
In Fig.~\ref{fig:size_v}, we show the results for $v_\mathrm{max}=5$ and 20 km s$^{-1}$.
As expected, small $v_\mathrm{max}$ predicts inefficient shattering because
of reduced grain--grain collision rates and small impact energies.
For $v_\mathrm{max}$ as large as 20 km s$^{-1}$,
shattering disrupts not only large grains but also small grains,
because the
grain velocity peaks at $a<0.1~\micron$ and it is as high as $\sim$10 km s$^{-1}$ in
a wide range of grain radii. Therefore, the overall grain abundance
also decreases (recall that we remove grains shattered into $a<3$ \AA).
This also means that shattering does not act as a production mechanism
of small $\sim 0.01~\micron$) grains, which affect the UV reddening,
if the turbulence velocity is too high.

\subsection{Evolution of extinction}

\begin{figure}
\includegraphics[width=0.48\textwidth]{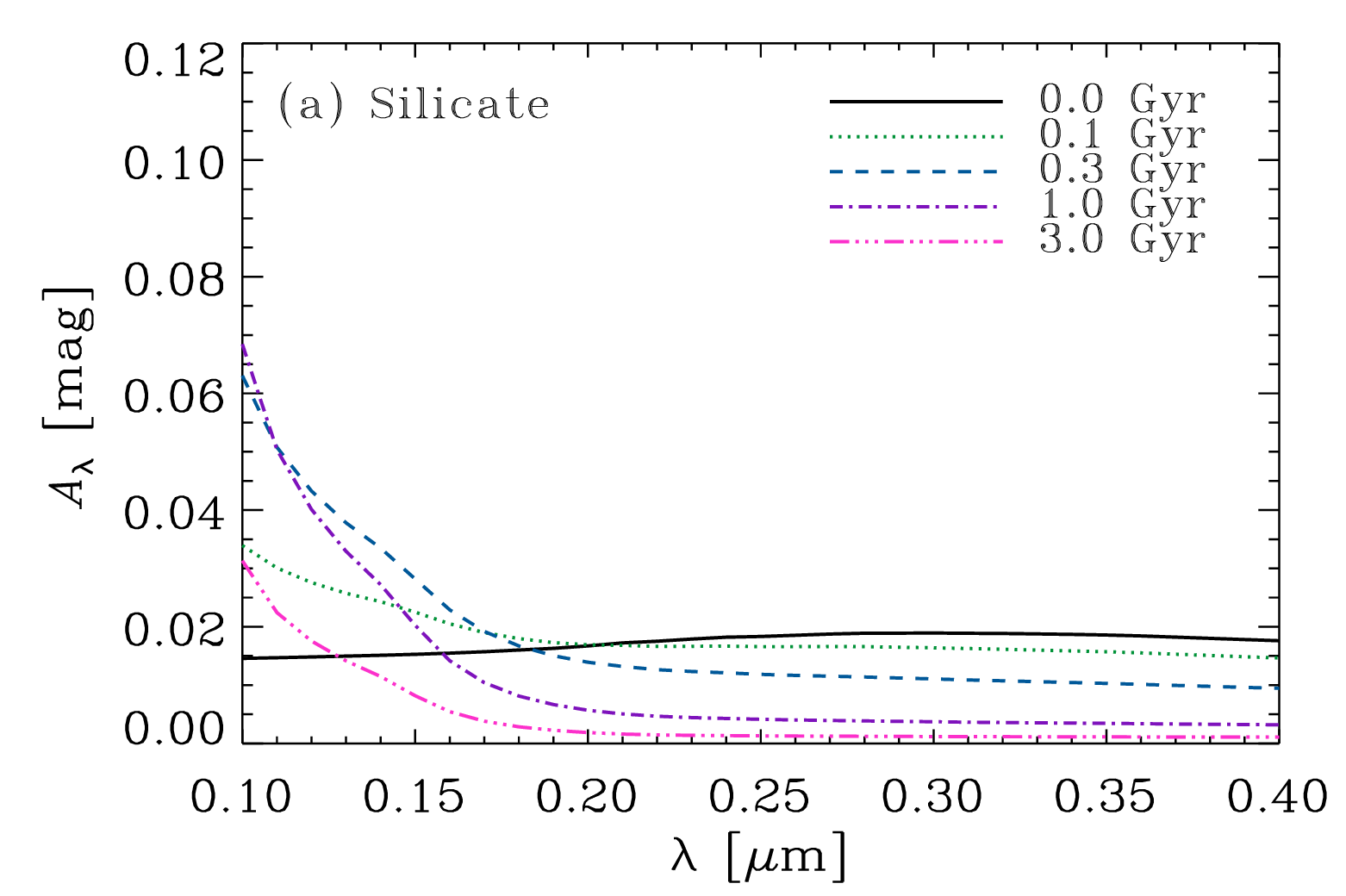}
\includegraphics[width=0.48\textwidth]{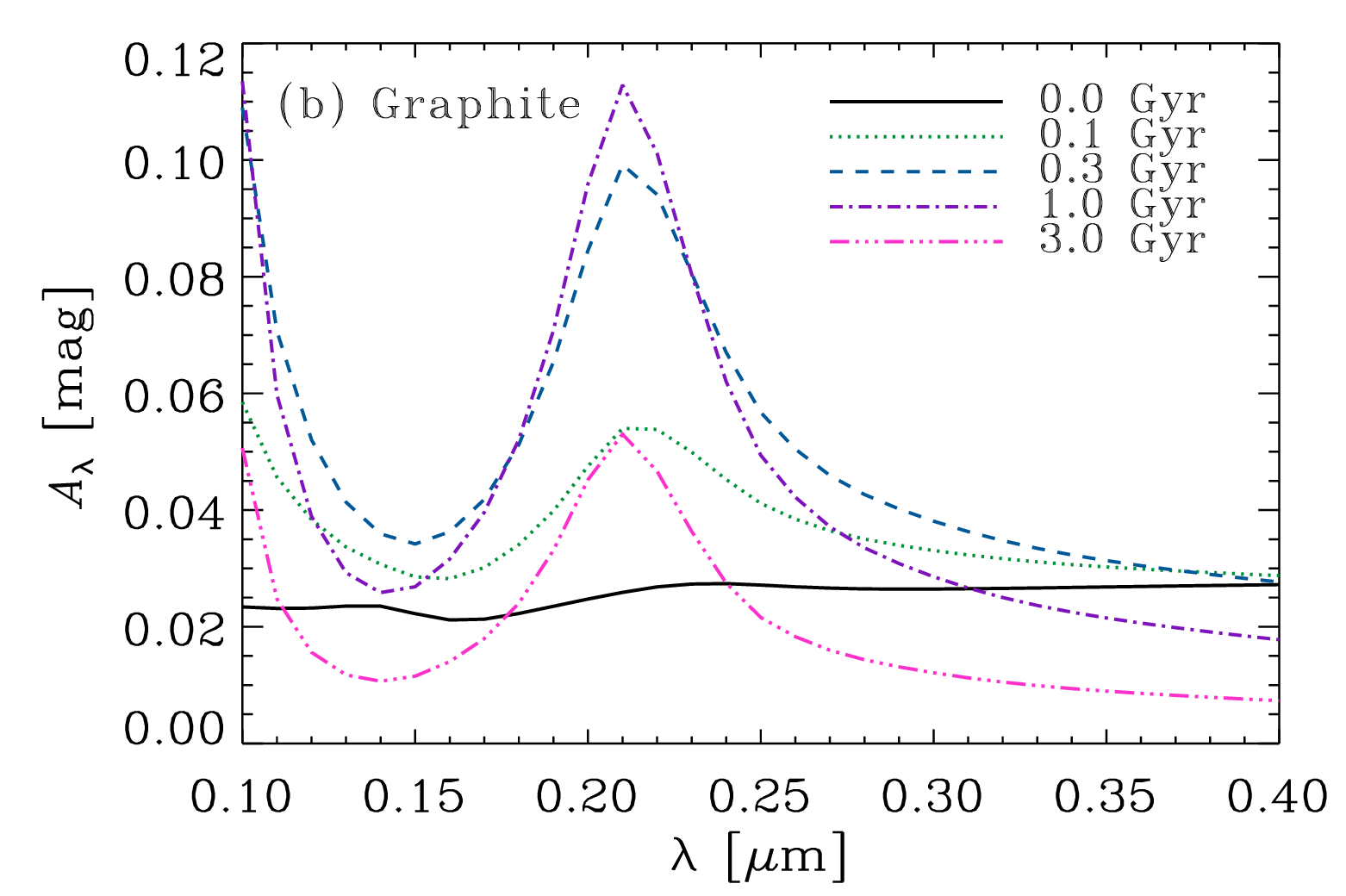}
\includegraphics[width=0.48\textwidth]{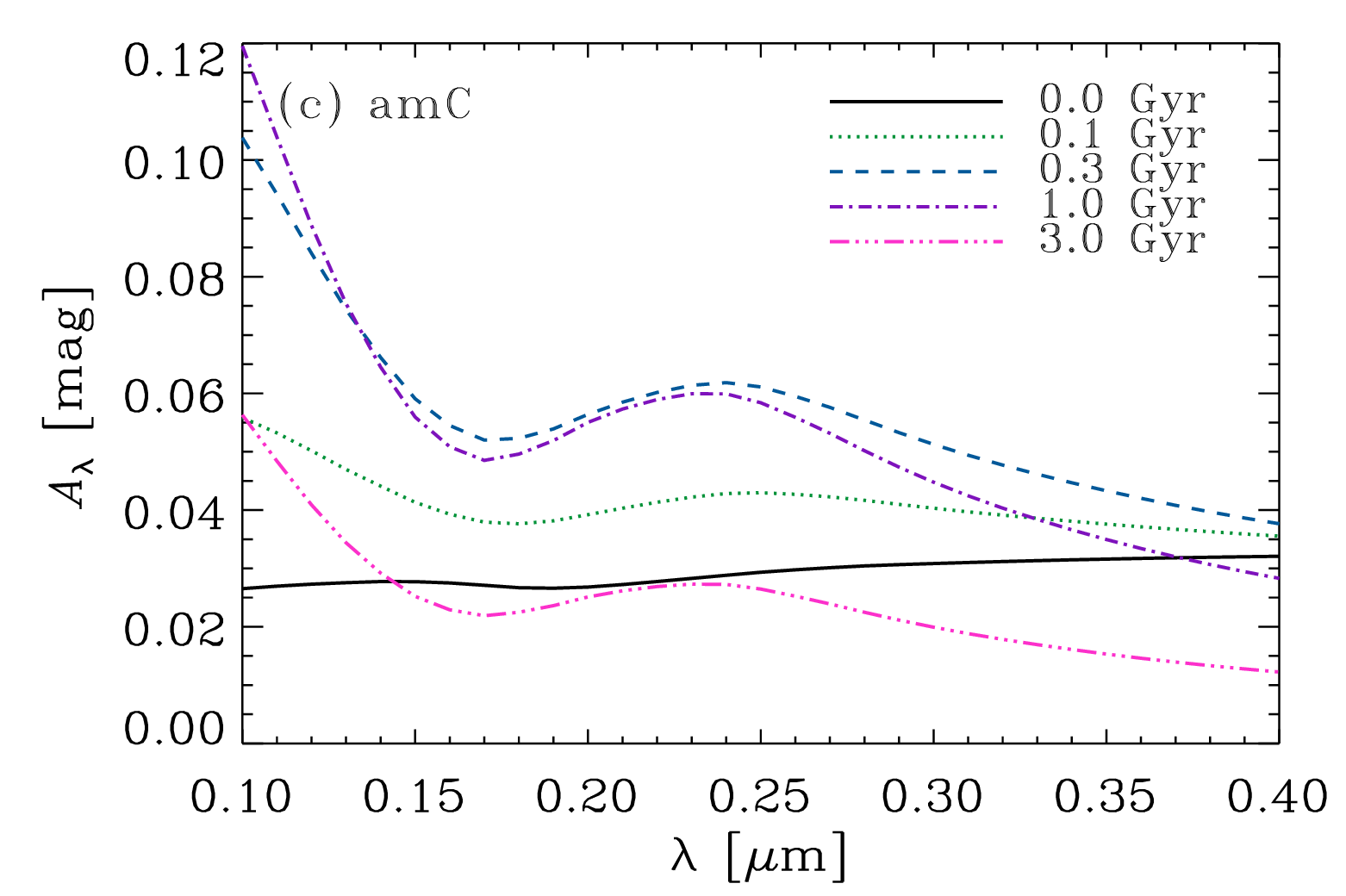}
\caption{Wavelength dependence of extinction ($A_\lambda$)
for (a) silicate, (b) graphite, and (c) amC.
The solid, dotted, dashed, dot--dashed, and triple-dot--dashed lines show the
results at $t=0$ (initial), 0.1, 0.3, 1, and 3 Gyr, respectively.
We adopt $N_\mathrm{H}=10^{19.5}$ cm$^{-3}$, but the resulting extinction
is simply proportional to $N_\mathrm{H}$.
\label{fig:ext}}
\end{figure}

Before examining the reddening curves, we show the evolution of extinction, $A_\lambda$,
with the fiducial parameter values.
In Fig.\ \ref{fig:ext}, we show the time variation of $A_\lambda$ (the extinction as a function
of rest-frame wavelength) for silicate, graphite, and amC.
We observe that the initial extinction has a flat dependence on the the wavelength
for all the grain species. Thus, there is no significant reddening in the initial condition,
which confirms that the grain size distribution dominated by large grains cannot
explain the reddening observed for Mg \textsc{ii} absorbers.
As small grains are produced by shattering, the extinction
rises steeply at short wavelengths up to $t\sim 0.3$--1 Gyr. After that, the extinction
decreases at almost all wavelengths because the loss from the lower
grain radius boundary becomes significant.
This grain loss is negligible up to $t=0.3$ Gyr, and the mass fraction of lost grains becomes
25 per cent at $t=1$ Gyr and 66 per cent at $t=3$ Gyr in the fiducial case.
Thus, the extinction at short wavelengths
is the most enhanced at $t\sim 0.3$--1 Gyr,
which can be regarded as a time-scale of extinction curve steepening.

The wavelength dependence of $A_\lambda$ is sensitive to the grain species.
Graphite has a prominent bump at 2175 \AA. This could help explaining the reddening
from optical wavelengths to $\lambda\sim 0.2~\micron$.  Both silicate and amC have
rising reddening towards shorter wavelengths except for the small bump of amC around
$\lambda\sim 0.25~\micron$. Thus, silicate and/or amC could be responsible for
rising extinction towards far-UV wavelengths.

The evolution of extinction shown here for the fiducial  case gives a basis
on which we interpret the evolution of reddening curve.
With $n_\mathrm{H}=10^{-1}$~cm$^{-3}$, the variation of the other parameters
broadly changes the time-scale of steepening the UV extinction
but the maximum steepness of the extinction curve does not exceed the above
fiducial case significantly.
For $n_\mathrm{H}=10^{-2}$ cm$^{-3}$, the wavelength dependence of $A_\lambda$
remains flat because of inefficient shattering. We mention the parameter dependence
again when we discuss the reddening in Section \ref{subsec:ext_result}.

\subsection{Reddening curves}\label{subsec:ext_result}

\begin{figure}
\includegraphics[width=0.48\textwidth]{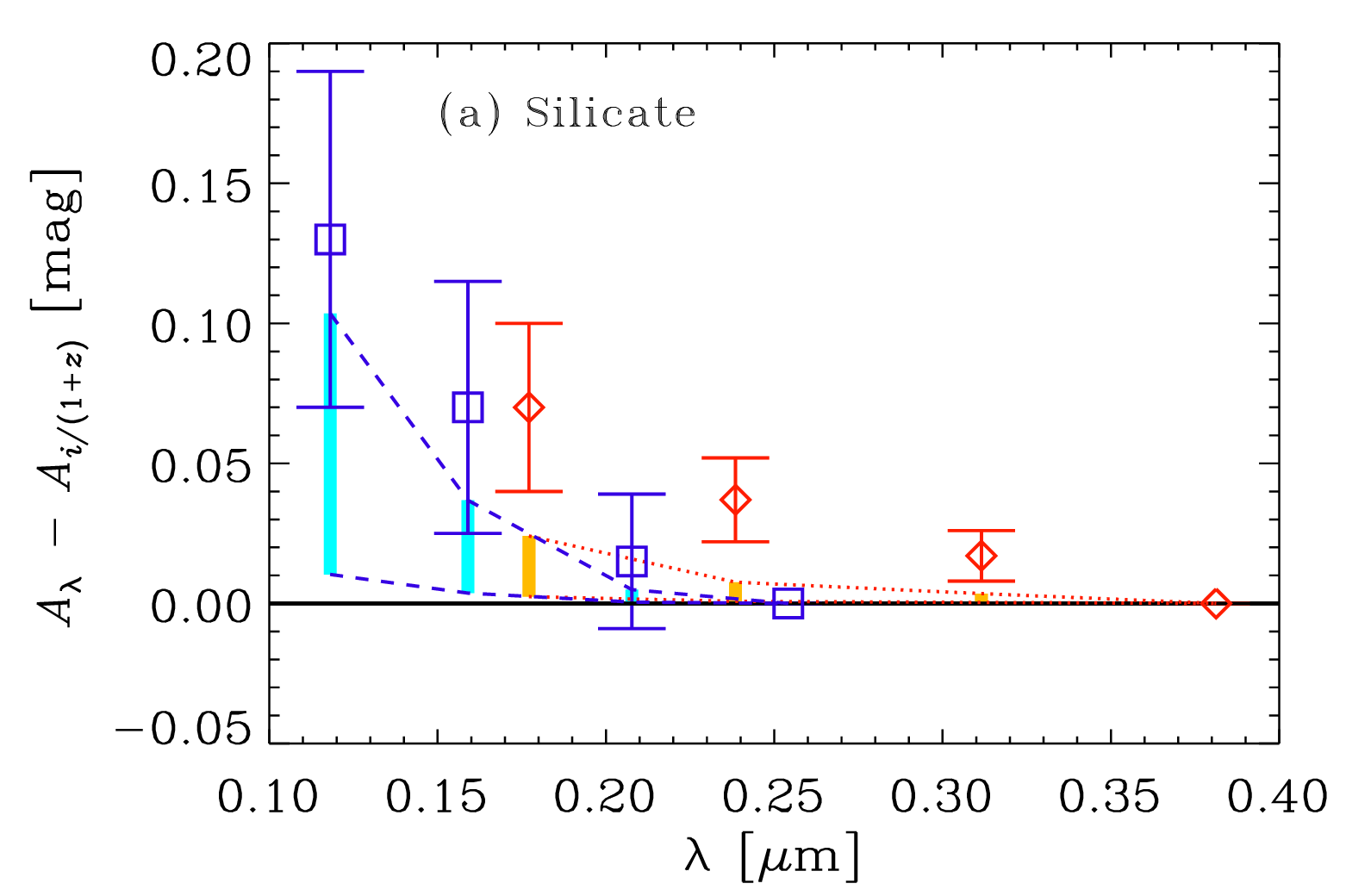}
\includegraphics[width=0.48\textwidth]{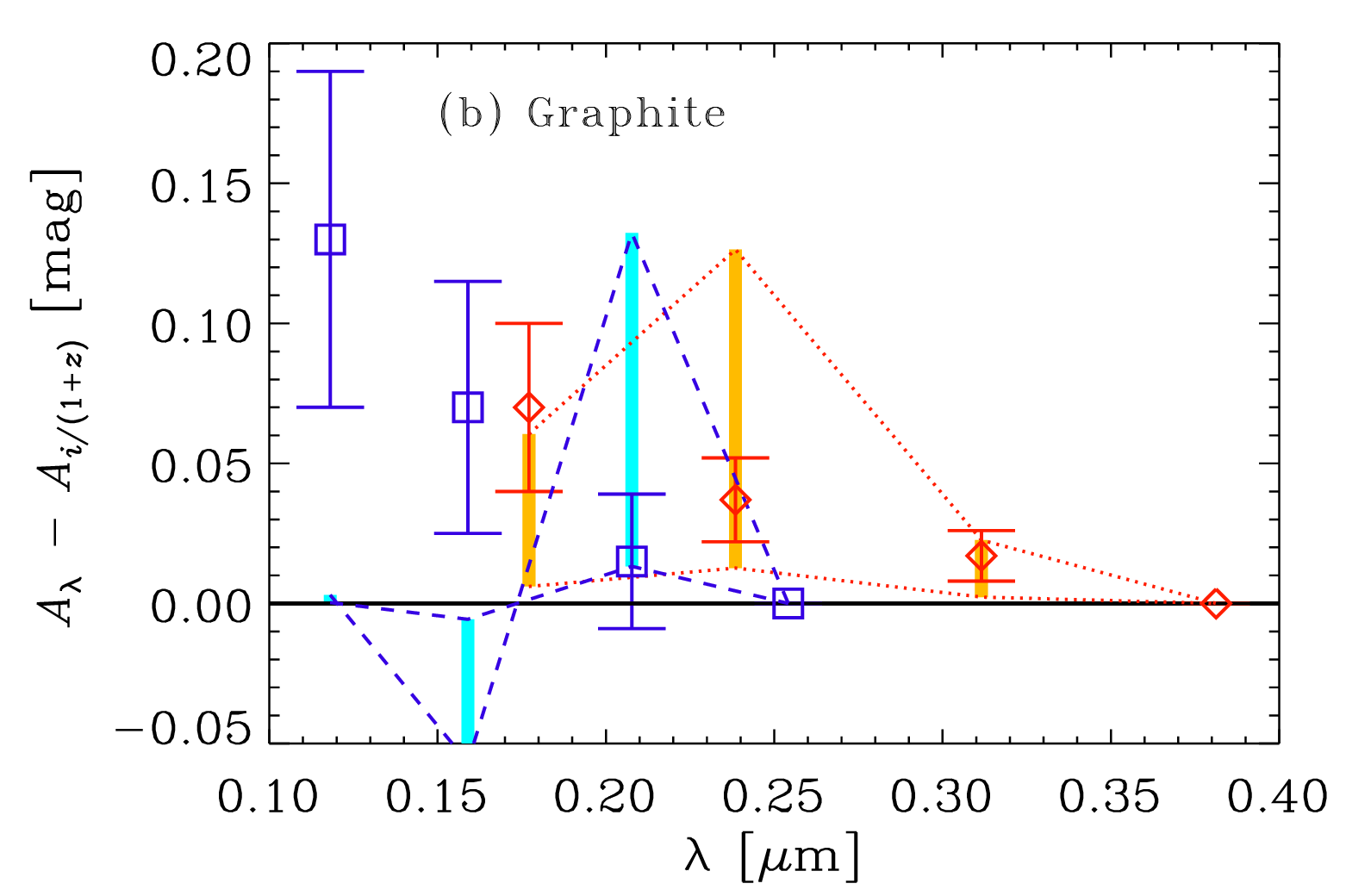}
\includegraphics[width=0.48\textwidth]{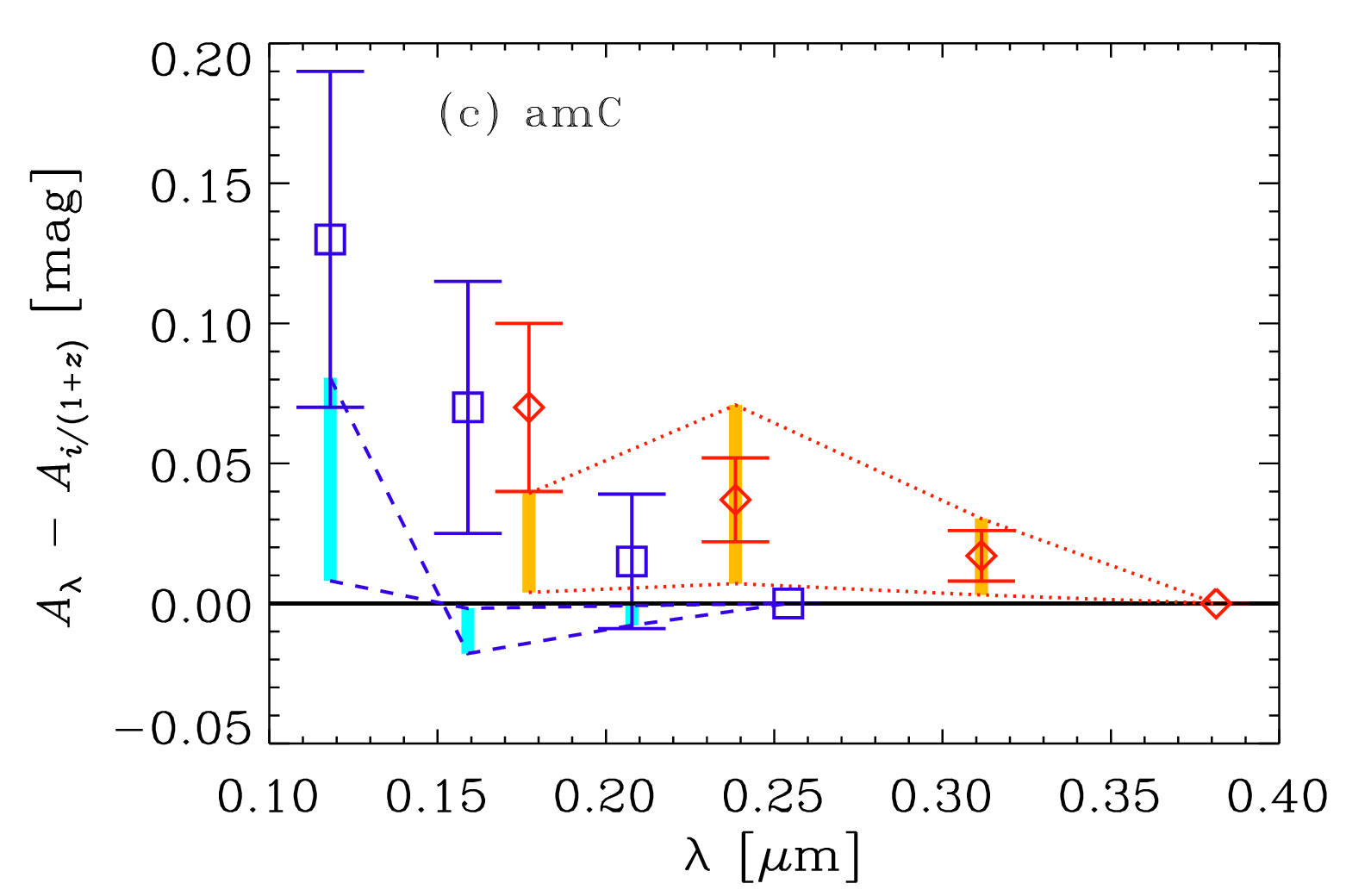}
\caption{Reddening curves (i.e.\ extinction relative to the $i$-band extinction)
for (a) silicate, (b) graphite, and (c) amC in the rest frame at $z=1$ and 2.
We adopt the grain size distribution for the fiducial case at $t=3\times 10^8$ yr.
The orange and light blue bars (connected by the lines in the same colour)
show the theoretically predicted ranges of reddening
in the SDSS $u$, $g$, and $r$ bands at $z=1$ and 2,
respectively. The range of each bar
corresponds to $N_\mathrm{H}=10^{19}$--$10^{20}$ cm$^{-2}$.
The red diamonds and blue squares with error bars show the observational data of
Mg\,\textsc{ii} absorbers at $z=1$ and 2, respectively, taken from MF12, with the errors
expanded by a factor of 3 for a conservative comparison.
\label{fig:reddening}}
\end{figure}

Based on the grain size distributions shown above, we calculate the reddening
curves by the method explained in Section~\ref{subsec:ext_method}.
Recall that the reddening curve
is the difference between the extinction at rest-frame wavelength $\lambda$
and that in the $i$-band in the observer's frame
[rest-frame wavelengths
$i/(1+z)=0.38$ and 0.25~$\micron$ at $z=1$ and 2, respectively];
that is, $A_\lambda -A_{i/(1+z)}$ as a function of $\lambda$.
Recall that we adopt the observational data taken from MF12. For a
conservative comparison, the errors in MF12 are expanded by a factor of 3.
We use the grain size distribution at $t=0.3$~Gyr.
After $t=1$ Gyr, the grain abundance starts to decrease, so that
the reddening is simply reduced by the effect of grain mass loss. Moreover, the
probable lifetime of the IGM clumps is $\lesssim\mbox{a few}\times 10^8$ yr
(Introduction). At $t\lesssim 0.1$ Gyr,
the grain size distribution does not develop enough to steepen the reddening curve.
Thus, we here judge if the steepening of reddening curve occurs or not by showing the
results at $t=0.3$ Gyr. Considering the uncertainties in $N_\mathrm{H}$, we show the
range of reddening expected for $N_\mathrm{H}=10^{19}$--$10^{20}$ cm$^{-2}$
following HL20.

In Fig.~\ref{fig:reddening}, we present the reddening curves corresponding to
the grain size distributions shown in Fig.~\ref{fig:size} (the fiducial case) at $t=0.3$~Gyr
for silicate, graphite, and amC.
{To avoid overlaying reddening curves at two redshifts,
we show the reddening values only at the wavelengths where observational data are available,
presenting the range of the reddening corresponding to
$N_\mathrm{H}=10^{19}$--$10^{20}$~cm$^{-2}$.
Naturally, the continuous shape of reddening curve for $\lambda$ preserves that of extinction
curve shown in Fig.\ \ref{fig:ext}.}
As we observe in Fig.~\ref{fig:reddening}, significant reddening emerges, but the details
depend on the grain species. For silicate, the predicted reddening is almost comparable to
the observational data at $z=2$, but is significantly smaller than the observations
at $z=1$. The difference in the calculated reddening curves between $z=1$ and 2
is simply due to the different reference wavelengths [$i/(1+z)$],
since we use the same grain size distribution for both redshifts.
Graphite, in contrast, reproduces the data points at $z=1$ better than
at $z=2$. The enhancement at $\lambda\sim 0.20$--0.25~$\micron$ is
due to the 2175~\AA\ bump
caused by small graphite grains. The other carbonaceous material, amC, reproduces the data
at $z=1$ except for the shortest wavelength, but not those at $z=2$.
Note that the reddening can be negative because $A_\lambda$ is not a monotonic function
of $\lambda$, especially for the reddening curve at $z=2$, whose zero-point is set
near the 2175~\AA\ bump of graphite and the
small 0.25 $\micron$ bump of amC.

Overall, compared with the initial extremely small reddening ($\lesssim 10^{-3}$),
the reddening curve is steepened by shattering. It is interesting to point out that
silicate and carbonaceous dust explain the steepening in a complementary way
in the sense that
the former and the latter species better explain the reddening at $z=1$ and 2,
respectively (see also Section~\ref{subsec:mixture}).
This implies that a mixture of silicate and carbonaceous dust {or different
dust materials at different redshifts better
explain the observed reddening curves}.

Now we discuss the dependence on the parameters,
whose effects on the grain size distribution have already been examined
in Section~\ref{subsec:size_result}.
As shown above, for a lower value of $n_\mathrm{H}=10^{-2}$ cm$^{-3}$,
the efficiency of small-grain production is lower. In this case, we confirm that
the reddening is $\lesssim 10^{-3}$. Thus, local density enhancement to
$n_\mathrm{H}\sim 10^{-1}$ cm$^{-3}$ as observationally
suggested by \citet{Lan:2017aa} is necessary
to explain the reddening in the CGM.

In our model, the fiducial case at $t\sim 0.3$~Gyr gives the largest reddening.
Other cases with varied parameters are described as follows
(we only change the parameter of interest with the others fixed to the fiducial values
as we did above).
In the case of $L_\mathrm{max}=30$~pc, the reddening at $t=0.3$~Gyr is roughly half
of that for the fiducial case. The reddening remains low in this case, reflecting less
efficient shattering than in the fiducial case (Section \ref{subsec:size_result}).
The reddening
reaches to a level comparable to that of the fiducial case at $t\sim 1$~Gyr in the
case of $L_\mathrm{max}=300$ pc, while its level is half at $t=0.3$ Gyr.
Recall that the small-grain production is the most efficient
around the fiducial value, $L_\mathrm{max}=100$ pc, since the grain velocity
as a function of grain radius for $L_\mathrm{max}=100$ pc peaks
around $a\sim 0.1~\micron$, coinciding with the {characteristic grain radius $a_0$}
of the initial grain size distribution. As mentioned above, the same value of
$T_\mathrm{gas}L_\mathrm{max}^2$ produces the same grain size distribution.
This means that, under a fixed $L_\mathrm{max}=100$ pc,
the case with $T_\mathrm{max}=10^4$ K
is optimum in causing reddening.

The grain size distribution also depends on $v_\mathrm{max}$. As shown
in Fig.\ \ref{fig:vel}(c), the grain velocities at all radii monotonically increases
as $v_\mathrm{max}$ increases. Thus, the reddening curve is steepened at an earlier stage
for larger $v_\mathrm{max}$. A comparable reddening curve for $v_\mathrm{max}=10$ km s$^{-1}$
at $t=0.3$ Gyr is reproduced with $v_\mathrm{max}=20$ km s$^{-1}$ at $t=0.1$ Gyr and
with $v_\mathrm{max}=5$ km s$^{-1}$ at $t=1$ Gyr. The reddening declines after these times
because of the loss of small grains from the lower grain radius boundary.
Thus, the maximum turbulence velocity determines
the overall time-scale on which shattering steepens the reddening curve.

We note that the assumed values of $N_\mathrm{H}$ and $\mathcal{D}_0$ are uncertain.
Since the resulting reddening is proportional to $N_\mathrm{H}\mathcal{D}_0$,
a factor 2 increase in $N_\mathrm{H}\mathcal{D}_0$ would easily explain the observed
level of reddening curves, although we do not fine-tune these parameters further.

{The reddening is also affected by the choice of the smallest grain radius, which
we assume to be 3 \AA. Some authors adopted larger values for the minimum grain radius
in modelling the grain size distribution
\citep[e.g.][]{Silvia:2010aa,Kirchschlager:2019aa,Hoang:2021aa}. As argued by
\citet{Nozawa:2013aa}, the smallest grain radius only has a minor influence on the extinction curves.
In our case, if we simply eliminate the grains with $a<10$ \AA\ (without keeping the dust mass),
the reduction of reddening at the shortest wavelength for silicate is 15 per cent, while
that at the bumps of carbonaceous grains is 20 per cent.}

\section{Discussion}\label{sec:discussion}

\subsection{Mixture of grain species}\label{subsec:mixture}

In the above, we have shown that steepening does occur by shattering associated with
cool clumps in the CGM, concluding that shattering could be a cause of
the reddening observed for Mg\,\textsc{ii} absorbers.
However, the degree of reddening depends on the grain species, and it seems to be
difficult to explain the reddening data at both $z=1$ and 2 with a single grain species.
Interestingly, silicate and carbonaceous dust are supplementary in the sense that the former species
explains the reddening at $z=1$ relatively well while the latter is more favoured at $z=2$.
In reality, it is more likely that the IGM dust is composed of multiple species.
Thus, it is intriguing to investigate if a mixture of
silicate and carbonaceous species explains the reddenings at $z=1$ and 2 at the same time.

\begin{figure}
\includegraphics[width=0.48\textwidth]{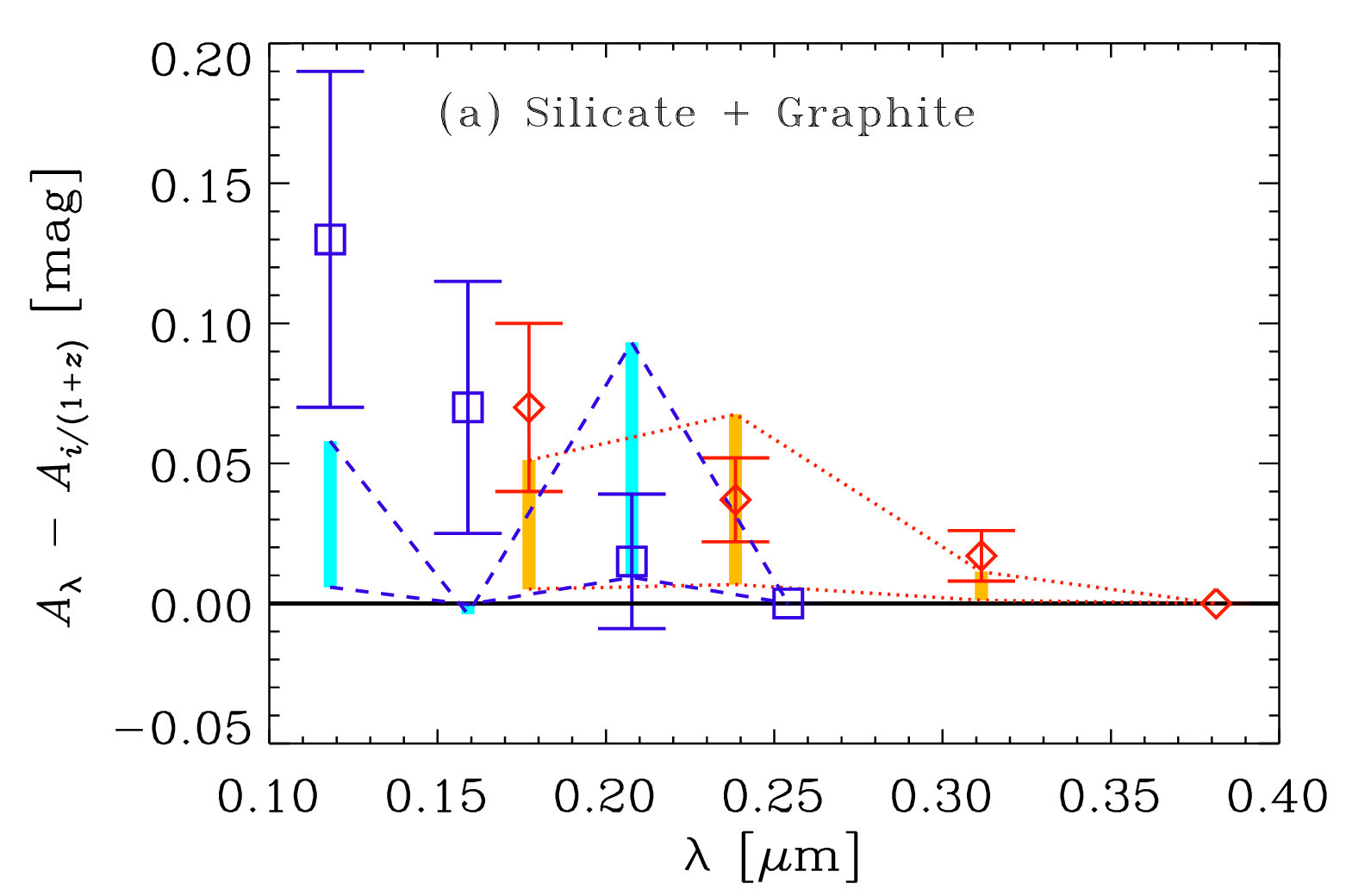}
\includegraphics[width=0.48\textwidth]{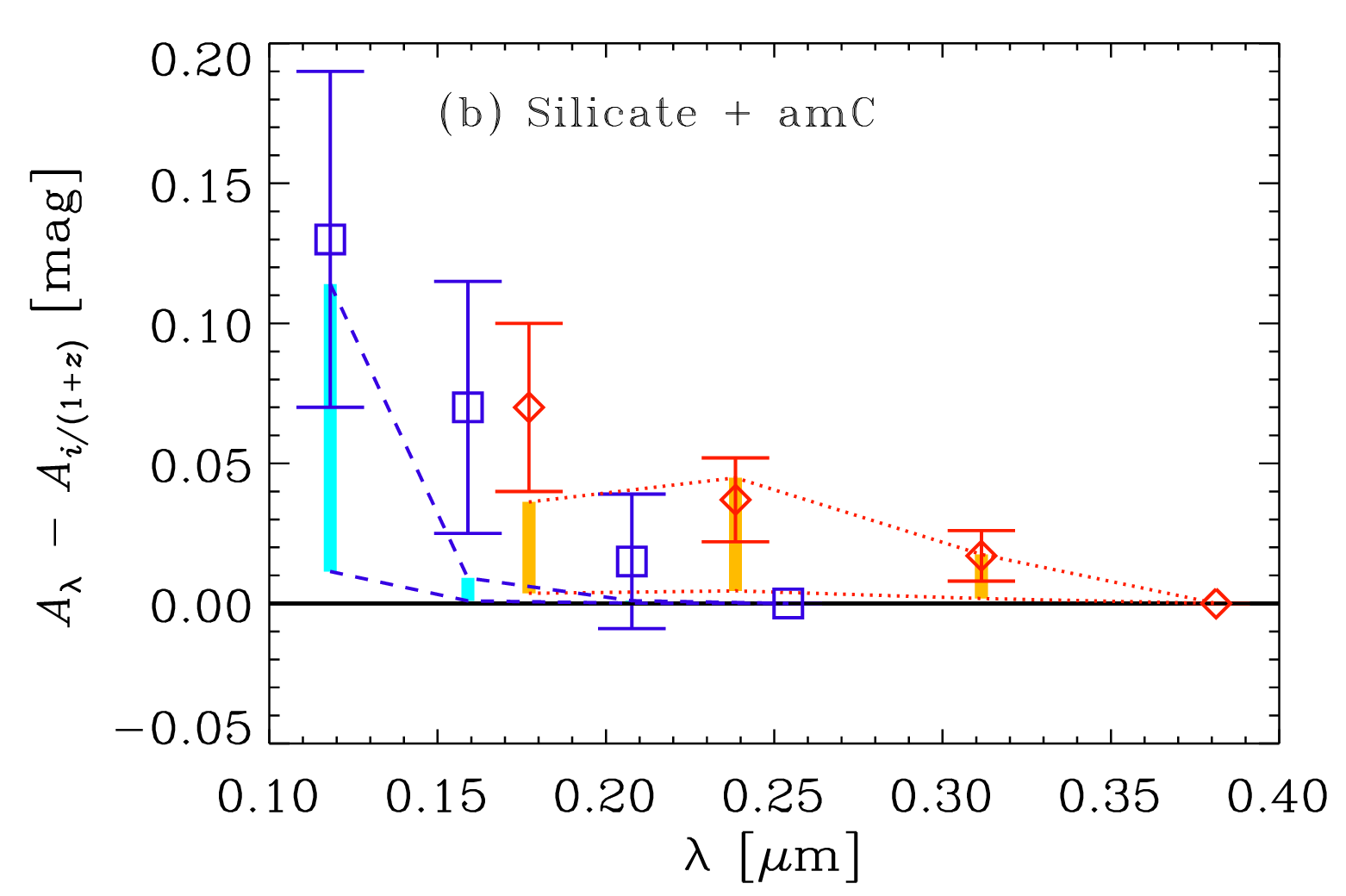}
\caption{Same as Fig.\ \ref{fig:reddening} but for the mixture of silicate and carbonaceous dust
with a mass ratio of 0.54:0.46.
Panels (a) and (b) show the mixtures of silicate and graphite, and of silicate and amC, respectively.
\label{fig:reddening_mixed}}
\end{figure}

Here we investigate the mixture of silicate and graphite as an example of multiple-species mixture.
We assume the mass ratio of silicate to graphite to be 0.54:0.46 \citep{Hirashita:2019aa}
{and add the extinctions of the two species at $t=0.3$ Gyr with the above weight.
This mass ratio is valid for the
Milky Way extinction curve \citep{Hirashita:2019aa}, and}
the almost half--half ratio is also useful to
maximize the effect of the mixture.
In Fig.\ \ref{fig:reddening_mixed}(a),
we show the results. As expected, the resulting reddening is significant
at both $z=1$ and
2, although the reddening data at the shortest wavelengths are better explained by
the single-species (silicate) case. However, we emphasize that the mixture makes
the behaviour of reddening
along the wavelength axis more moderate in the sense that the specific feature of a certain
single species becomes less prominent. We note that graphite-dominated dust is
not favoured by
the extinction curve derived for QSO absorbers
by \citet{York:2006aa}, which does not show any significant 2175 \AA\ bump.

We also examine the mixture of silicate and amC in Fig.\ \ref{fig:reddening_mixed}(b), where
we adopt the same silicate-to-carbonaceous dust mass ratio as above.
We confirm that the mixture produces significant reddening at both $z=1$ and 2.
Which of graphite and amC better explains the reddening data
depends on the wavelength; thus, we are not able to conclude which dust species is
favourable.

We also point out that the ratio between silicate and
carbonaceous dust could depend on the redshift. Since clarifying the redshift dependence
theoretically is beyond the scope of our formulation, we leave the evolution of grain species in
the CGM for a future work.

\subsection{Possible solutions to the underproducing tendencies}

In the above results, both silicate and carbonaceous species tend to underproduce the
observed reddening curves, especially at short wavelengths. As mentioned above,
the extinction is proportional to
the product of $N_\mathrm{H}$ and $\mathcal{D}$, both of which are uncertain.
We already assumed $\mathcal{D}_0=0.006$ (the initial value of $\mathcal{D}$),
which is appropriate in solar metallicity
environments. If Mg \textsc{ii} absorbers trace metal-enriched gas originating from
the central galaxy, it is possible that the dust-to-gas ratio is higher. If the dust-to-gas
ratio is, for example, twice higher than assumed in this paper, the reddening curves
are almost consistent with the observational data.
Adopting a fixed range of $N_\mathrm{H}$ might be too simplistic;
indeed, there is probably a larger
dispersion in $N_\mathrm{H}$ for Mg \textsc{ii} absorbers with an indication of
redshift evolution \citep{Lan:2017aa}. Thus, more sophisticate modelling may be necessary;
for example, the detailed distribution function of $N_\mathrm{H}$
dependent on $z$ would
be worth modelling. However, we emphasize that, even with our simple models,
shattering produces significant reddening, which is consistent with the observational
data within a factor of $\sim$2.

\citet{Lan:2017aa} showed that the column density
of Mg \textsc{ii} absorbers becomes higher with redshift as
$N_\mathrm{H}\propto (1+z)^{1.9}$.
Thus, we expect that $N_\mathrm{H}$ is $(3/2)^{1.9}\simeq 2.2$ times higher
at $z=2$ than at $z=1$.
This could make it easier to explain the data points at $z\sim 2$.
We should keep in mind, however, that $\mathcal{D}$ may also be different between
$z=1$ and 2 \citep{Peroux:2020aa}.

The treatment of shattered fragments could be uncertain, especially for extremely small
grains. Applying the bulk material properties to grains
near the lower grain radius boundary ($a\sim 3$ \AA) may not be valid.
For example, if such small grains are not efficiently disrupted by shattering,
the grains are not lost by shattering as efficiently as we calculated in this paper.
However, it is difficult to introduce this kind of effect
since we do not have knowledge on the disruptive properties at such small grain radii.
Although it is difficult to fully address this issue, we make an attempt of effectively examining the
effect of grain mass loss in shattering in the following.

\begin{figure}
\includegraphics[width=0.48\textwidth]{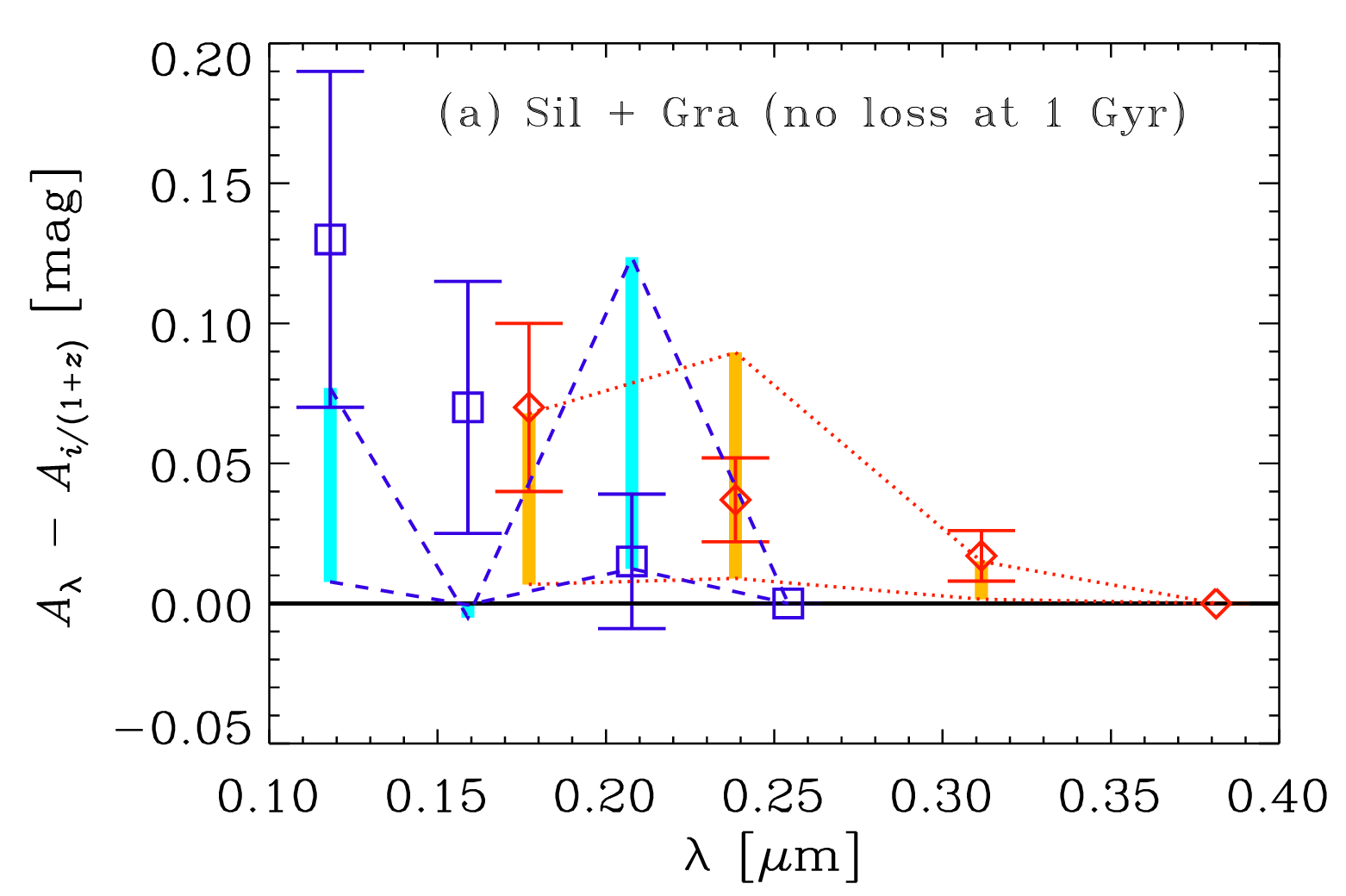}
\includegraphics[width=0.48\textwidth]{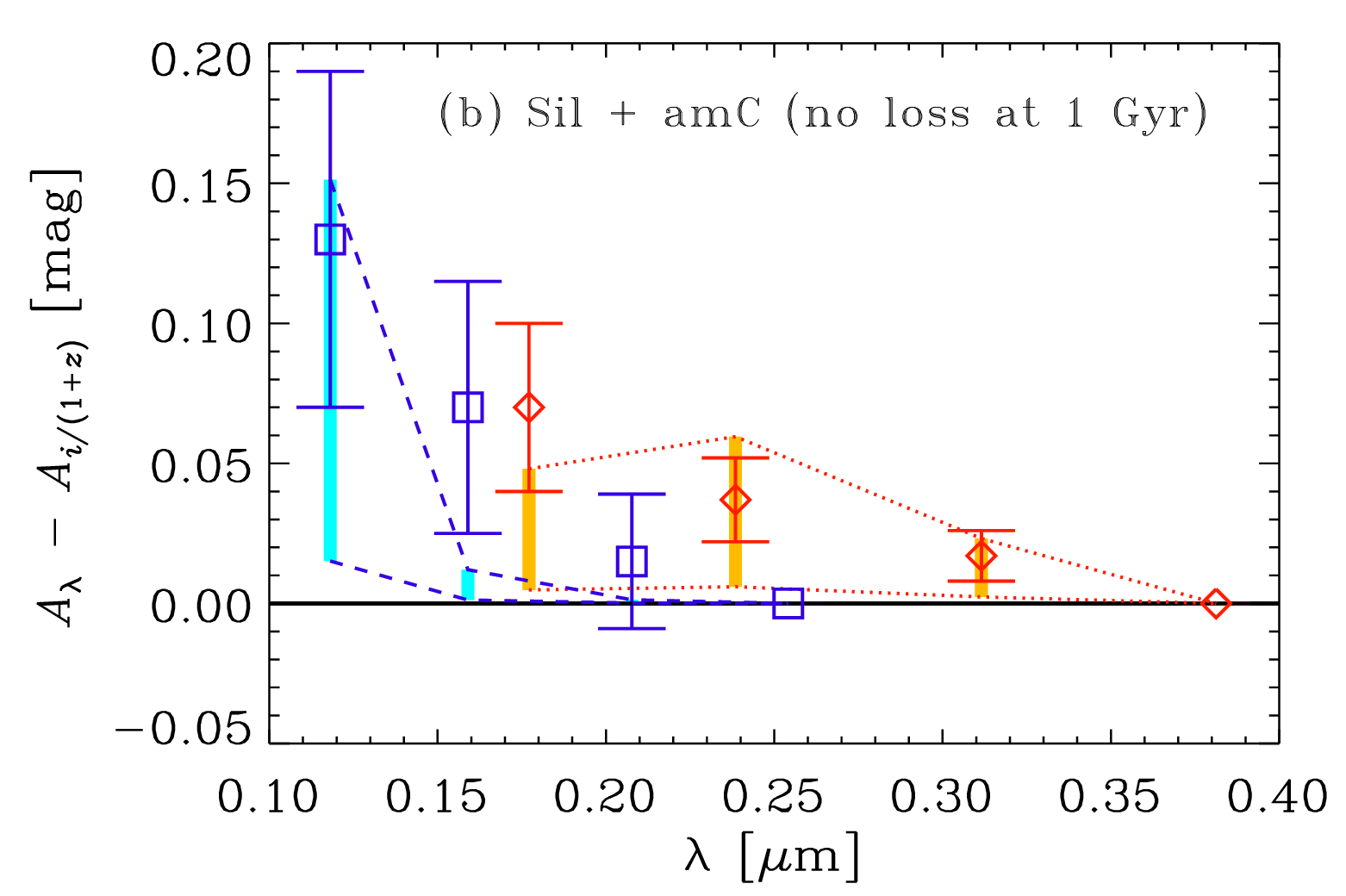}
\caption{Same as Fig.\ \ref{fig:reddening_mixed} but with the correction
for dust mass loss using the renormalization described in the text
(to supplement the lost dust).
\label{fig:reddening_noloss}}
\end{figure}

In the above, the dust abundance (dust-to-gas ratio $\mathcal{D}$) decreases
as a function of time, and the decreasing effect is prominent at $t\gtrsim 1$~Gyr
in the fiducial case. To mimic the case where small grains are not efficiently shattered,
we adopt $\mathcal{D}=0.006$ (the initial value) instead of the value at each age,
in calculating the extinction (equation~\ref{eq:A_lambda}).
This effectively cancels out the effect of grain mass loss by artificially adjusting
$\mathcal{D}$. This `correction' for the dust mass loss is important only at
$t\gtrsim 1$ Gyr. Thus, we only show in Fig.\ \ref{fig:reddening_noloss}
the result at $t=1$ Gyr for the same mixed cases
as in Fig.\ \ref{fig:reddening_mixed} (at $t=3$ Gyr, the correction factor is too large
to make a meaningful prediction). Comparing Fig.\ \ref{fig:reddening_noloss}
with Fig.\ \ref{fig:reddening_mixed}, we find that the reddening is enhanced.
Similar enhancement is also expected for reddening curves of single species.
Thus, a longer shattering duration without grain mass loss may be a solution for the
underprediction of reddening.

\subsection{Prospects for further modelling}\label{subsec:prospect}

In this paper, we have focused on shattering in the CGM. Here we
discuss our results in the context of overall dust evolution in the CGM.

\citet{Aoyama:2018aa}, using a cosmological simulation, showed that the dust-to-metal ratio
decreases with the distance from the galaxy centre in the CGM. This is due to sputtering in the
hot gas. However, in their simulation, the formation of
cool clumps in the CGM cannot be spatially resolved. The density structures in the CGM
calculated by cosmological simulations still depend on the spatial resolution
\citep[e.g.][]{vandeVoort:2019aa}, and it is in general extremely difficult for cosmological simulations
to resolve a scale of a few tens of parsecs
unless we develop some dedicated high-resolution scheme \citep{Hummels:2019aa}.
Thus, it is necessary to somehow include shattering associated with cool clumps
by developing a sub-grid model. The formulation or the result of this paper could be used as
a sub-grid input model in cosmological simulations.

There are some possible effects of shattering in the CGM. First, since the grains become
smaller, they are more efficiently destroyed by sputtering once they are injected into the hot
($T_\mathrm{gas}\gtrsim 10^6$ K) gas.
Thus, the modification of grain size distribution by shattering could alter
the grain destruction rate by sputtering in the CGM.
Secondly, we expect that the grain size distribution is inhomogeneous in the CGM with more small grains
in cool clumps. Thus, if the dust observation is biased to the cool medium (such as Mg \textsc{ii}
absorbers), it selectively see regions where UV reddening curves are steepened.
Thirdly, the photoelectric heating rate of the CGM and IGM depends on the grain radius.
Small grains tend to be more efficient in photoelectric heating with a fixed dust mass abundance
\citep{Inoue:2003ab,Inoue:2004aa}. Thus, the small-grain production by shattering
is important for the heating rate in the CGM (and in the IGM if the grains are transported to a wide area
in the Universe).
It is interesting to examine these effects of small-grain production self-consistently
by including shattering in the CGM in
cosmological simulations, or to develop some analytical model that could include all relevant processes
for the enrichment and processing of dust in the CGM.

\section{Conclusion}\label{sec:conclusion}

We investigate the small-grain production in the CGM through shattering,
which could occur efficiently in turbulent media with a high enough gas density.
Cool ($T_\mathrm{gas}\sim 10^4$~K) clumps traced by Mg\,\textsc{ii} absorbers are
possible sites that have favourable conditions for shattering.
We calculate the evolution of grain size distribution
by shattering in turbulent cool gas and examine if small grains are efficiently produced on
a reasonable time-scale (i.e.\ within the typical lifetime of the cool gas
$\sim\mbox{a few}\times 10^8$ yr).

We solve the shattering equation in a condition relevant for cool clumps in the CGM.
For the initial condition, we adopt a lognormal grain size distribution with
{characteristic grain radius $a_0\sim 0.1~\micron$}.
This is because previous studies suggested that grains transported from the central galaxy to the CGM
are dominated by such large grains. This initial condition also serves to examine
if shattering could produce a sufficient number of small grains within a reasonable time
even from the large-grain-dominated grain size distribution.
The grain motion is assumed to be induced by turbulence whose
maximum size is determined roughly by the clump size ($L_\mathrm{max}\sim 100$ pc).
It is also expected that the maximum velocity of
turbulent motion is of the order of $v_\mathrm{max}\sim 10$ km s$^{-1}$,
which is comparable to the sound speed of the cool gas.
The initial dust-to-gas ratio is fixed to $\mathcal{D}_0=0.006$
as derived from observations (noting that the time-scale simply scales as $\mathcal{D}_0^{-1}$).

We find that small-grain production efficiently occurs in the fiducial parameter set
($n_\mathrm{H}=0.1$ cm$^{-3}$, $T_\mathrm{gas}=10^4$ K, $L_\mathrm{max}=100$ pc, and
$v_\mathrm{max}=10$ km s$^{-1}$) on a time-scale of a few $\times 10^8$ yr, comparable to the
lifetime of cool clumps. It also turns out
that the above values for $n_\mathrm{H}$, $T_\mathrm{gas}$ and
$L_\mathrm{max}$ are optimum for small-grain production in the sense that the peak of the grain
velocity as a function of grain radius is located {around the characteristic grain
radius $a_0\sim 0.1~\micron$}.
Note that the above fiducial values are chosen not for optimizing the shattering efficiency but
based on the values suggested from observations of Mg\,\textsc{ii} absorbers.

We further calculate reddening curves based on the above grain size distributions but using
various dust species. We find that
the reddening becomes significant at $t\sim\mbox{a few}\times 10^8$ yr
in the fiducial parameter values.
The time-scale of reddening is similar to the lifetime of cool clumps in the CGM. Therefore,
we conclude that shattering in the CGM is a probable reason for the observed reddening.
The reddening curves are sensitive to the grain species. Silicate explains the reddening observed at
$z=2$ while it underpredicts that at $z=1$.
In contrast, graphite and amC explain the reddening observed at $z=1$ more easily than that at $z=2$.
The difference between $z=1$ and 2 is due to the different rest-frame wavelengths,
$i/(1+z)$, used for the zero point
(note that our model applies the same grain size
distribution to both redshifts).
A mixture {(or redshift-dependent fraction)} of silicate and carbonaceous dust
is favoured to explain the reddenings at $z=1$ and 2
simultaneously.

There is still a tendency that our model underpredicts the reddening. Considering that
the predicted reddening level is proportional to the hydrogen column density and the
dust-to-gas ratio, the underprediction could be associated with the large uncertainties in
these quantities. Moreover, the treatment of shattering at very small grain radii
($a\lesssim 0.001~\micron$) may need improvement since we simply apply bulk material
properties. If such extremely
small grains are not efficiently shattered, they remain to contribute to the reddening.

Since shattering associated with cool clumps has proven to be important,
it is desirable to somehow include this process in dust evolution models in a cosmological volume.
For example, a sub-grid model could be developed for shattering to be included in a cosmological
hydrodynamic simulation. This kind of development will be worth tackling in a future work.

\section*{Acknowledgements}
 
{We are grateful to the anonymous referee for useful comments.}
HH thanks the Ministry of Science and Technology (MOST) for support through grant
MOST 107-2923-M-001-003-MY3 and MOST 108-2112-M-001-007-MY3, and the Academia Sinica
for Investigator Award AS-IA-109-M02.
TWL acknowledges support from NSF grant AST1911140 and the visitor support from
Institute of Astronomy and Astrophysics, Academia Sinica.

\section*{Data Availability}

Data related to this publication and its figures are available on request from
the corresponding author. 


\bibliographystyle{mnras}
\bibliography{/Users/hirashita/bibdata/hirashita}




\bsp	
\label{lastpage}
\end{document}